\documentstyle[12pt,epsf,axodraw]{article}

\newcommand{\mcemptyr}{\multicolumn{2}{c|}{\mbox{}}}

\def\0{\over } 
\def\1{\vec }     
\def\2{{1\over2}} 
\def\4{{1\over4}}            
\def\5{\bar }  
\def\6{\partial } 
\def\7#1{{#1}\llap{/}}                         
\def\8#1{{\textstyle{#1}}}
\def\9#1{{\bf {#1}}}                           
 
\def\llp{\hbox to 0pt{\hss/\hskip1.5pt}}
\def\llo{\hbox to 0.2pt{\hss /}} \def\llq{\hbox to 0pt{\hss/\hskip0.5pt}}
\def\so{\supset\hbox to 0pt{\hss $\displaystyle -$\hskip1pt}}

\def\<{\langle } \def\>{\rangle }

\def\bea{\begin{eqnarray}} \def\eea{\end{eqnarray}} 
\def\beann{\begin{eqnarray*}} \def\eeann{\end{eqnarray*}} 
\def\beq{\begin{equation}} \def\eeq{\end{equation}}  
\newcommand{\la}[1]{\label{#1}}
\newcommand{\be}{\begin{equation}}
\newcommand{\ee}{\end{equation}}
\newcommand{\ba}{\begin{eqnarray}}
\newcommand{\ea}{\end{eqnarray}}
\newcommand{\rmi}[1]{{\mbox{\scriptsize #1}}}
\newcommand{\nr}[1]{(\ref{#1})}
\newcommand{\tr}{{\rm Tr\,}}
\newcommand{\nn}{\nonumber \\ }
\newcommand{\fr}[2]{{\frac{#1}{#2}}}
 
\def\lsi{\raise0.3ex
\hbox{$<$\kern-0.75em\raise-1.1ex\hbox{$\sim$}}}
\def\gsi{\raise0.3ex
\hbox{$>$\kern-0.75em\raise-1.1ex\hbox{$\sim$}}}

\begin{document} 
\setlength{\baselineskip}{18pt}                                     
\thispagestyle{empty}
\begin{flushright}
{hep-lat/9711022\\ 
CERN-TH/97-309\\
HD-THEP-97-55\\
November 1997}
\end{flushright}
\vspace{5mm}
\begin{center}
{\Large \bf
 Gauge-invariant scalar and field strength \\
 correlators in 3d}\\ \vspace{15mm}
{\large M.~Laine$^{\rm a,}$\footnote{mikko.laine@cern.ch} 
and O.~Philipsen$^{\rm b,}$\footnote{o.philipsen@thphys.uni-heidelberg.de}
}\\ \vspace{10mm}
{\it 
$^{\rm a}$CERN/TH, CH-1211 Geneva 23, Switzerland}\\ \vspace{5mm}
{\it 
$^{\rm b}$Institut f\"ur Theoretische Physik, Philosophenweg 16, \\ 
D-69120 Heidelberg, Germany}

\end{center}
\vspace{2cm}

\begin{abstract}
\thispagestyle{empty}
\noindent
Gauge-invariant non-local scalar and field strength 
operators have been argued to
have significance, e.g., as a way to determine the behaviour of the
screened static potential at large distances, as order parameters for
confinement, as input parameters in models of confinement, and as
gauge-invariant definitions of light constituent masses in bound state
systems.  We measure such ``correlators" in
the 3d pure SU(2) and SU(2)+Higgs models on the lattice.  We extract the
corresponding mass parameters and discuss their scaling and physical
interpretation.  We find that the finite part of the 
$\overline{\mbox{\rm MS}}$ scheme 
mass measured from the field strength correlator is large, more than
half the glueball mass. We also determine the non-perturbative
contribution to the Debye mass in the 4d finite $T$ SU(2)
gauge theory with a method due to Arnold and Yaffe, 
finding $\delta m_D\approx 1.06(4)g^2T$.
\end{abstract} 
\setcounter{page}{0}

\newpage

\section{Introduction}

The strength and range of forces described by gauge theories 
are characterized by the potential of a static charge.
Non-Abelian pure gauge theories in three and four dimensions are 
known to be confining in the sense that the potential 
in the fundamental representation rises linearly with distance.
If matter fields
are added the static potential is modified:
the linear rise persists only up to a critical distance at 
which enough energy is stored in the flux tube to pair create matter. 
The matter particles will then be separated and form bound states 
with the sources, so that 
the rise of the static potential saturates at the 
energy of the two bound state systems. 
If one considers static sources in the 
adjoint representation instead, the potential again rises linearly, 
but is supposed to saturate even in the pure gauge theory 
due to the creation of a pair of gluons.

In this work, we 
investigate non-perturbatively operators related to 
the energy of the bound state systems 
created after the breaking of the flux tube 
in the pure SU(2) and SU(2)+Higgs models. 
The non-local gauge-invariant operators measured 
consist of sources
of a given charge coupled via a Wilson line in the same representation:
\ba
G_\phi(x,y) & \equiv &  
\fr12 \left\< \tr \phi^\dagger(x) U^{\rm fund}(x,y)\phi(y)\right\> , 
\la{GPhi}\\
G_{F,ijkl}(x,y) & \equiv & 
\left\< F^a_{ij}(x) U^{\rm adj}_{ab}(x,y)F^b_{kl}(y)\right\>. 
\la{GA}
\ea
The fundamental Wilson line here is $U^{\rm fund}(x,y) =  {\cal P} 
\exp({ig\int_x^y\! dx_i A_i^a T^a})$
and the adjoint Wilson line is 
$U^{\rm adj}_{ab}(x,y) = 
2 \tr T^a U^{\rm fund}(x,y)T^b [U^{\rm fund}(x,y)]^\dagger$,
where $T^a=\sigma^a/2$ and $\sigma^a$ are the Pauli matrices.

More specifically, 
the mass signal $M$ extracted from these operators has a number of
applications.
First, related to the discussion above, 
$M$ is naively 
expected to determine the asymptotic value to which the static 
potential saturates, $V(\infty)=2 M$ (see below). 
The most prominent way of calculating the static
potential is to extract it from the expectation value of large
Wilson loops measured in lattice Monte Carlo simulations.
In four dimensions (4d) this has been
done for the fundamental representation 
(\cite{bali} and references therein) as well as for the adjoint
representation (\cite{mic85} and references therein). 
Analogous calculations in three dimensions (3d)
may be found in \cite{ilg,pou97}. 
However, no hard evidence for the saturation of the potential could
be obtained in these calculations. One possible explanation is
that the string in a lattice simulation does not break because of a
potentially bad projection of the Wilson loop onto the hadronized 
final state. Furthermore, Wilson loop calculations 
are very expensive in computer time. However, 
assuming the 
validity of the above picture of flux tube breaking, one may
combine knowledge of the static potential at small distances
with its asymptotic value obtained from 
measuring $M$, to arrive
at a rough estimate of the critical distance where the potential
flattens off, i.e., the ``screening length".

Second, the asymptotic
behaviour of $G_\phi(x,y)$ constitutes an order parameter for
the phase of the theory: in a Coulomb phase, 
$G_\phi(0,y) \sim \exp(-m|y|)/|y|^{(d-1)/2}$, whereas  
in a confinement or Higgs phase, the behaviour should be purely
exponential~\cite{bri83}. 

Third, the parameters related to the operator
$G_{F,ijkl}(x,y)$ contain important non-perturbative information about
the QCD ground state~\cite{gromes} and may 
hence act as input parameters in models
of confinement, such as the stochastic 
vacuum model~\cite{do87,do95} (for recent
references see, e.g., [9--11]).  
This is of relevance also in 3d, since 3d theories
can be regarded as laboratories for studying the
qualitative features of confinement in QCD~\cite{qcd3,ptw97}. 
The advantage of the 3d theories is that they
are superrenormalizable and,
consequently, exhibit very good
scaling behaviour so that one can extrapolate results of 
lattice simulations to the continuum limit with an accuracy 
at the percent level~\cite{kaj96,ptw96}. 
One especially useful feature of the SU(2)+Higgs model is that
the presence of the Higgs doublet allows a smooth interpolation
between the non-perturbative confinement and the perturbative 
Higgs regimes~\cite{isthere?,ptw97}.
The physical significance of the 3d theories stems
from the fact that they constitute the high temperature effective theories 
of usual 4d theories in the 
framework of dimensional reduction \cite{old,dr}.

Fourth, in analogy with the 
heavy -- light quark system~\cite{ei88}, the mass
parameters extracted from $G_\phi(x,y)$ and $G_{F,ijkl}(x,y)$
can be viewed as determining the masses of the light dynamical constituents  
in meson-like bound states. This aspect can be 
of relevance, e.g., 
for the constituent models proposed to apply in the
(symmetric) confinement phase
of the 3d SU(2)+Higgs model~\cite{do95,bp96}. In this context, the 
Wilson line operators have also been conjectured~\cite{bp96} 
to explain the value of the propagator
mass obtained from a simulation in a fixed Landau gauge \cite{kar}.

Finally, related to the determination of the light constituent
mass of a bound state, the operator $G_{F,ijkl}(x,y)$
in the pure 3d SU(N) theory can be used to measure the
leading non-perturbative contribution to the finite temperature 
Debye mass in 4d SU(N) QCD~\cite{ay}. The connection to 4d physics
is again via dimensional reduction.

The purpose of this paper is to measure the operators in 
eqs.~\nr{GPhi}, \nr{GA} in 3d (in 4d, 
these operators have been studied on the lattice
in~\cite{mic85,ev86}). We improve upon the 
measurements of $G_{F,ijkl}(x,y)$ performed
for pure 3d SU(2) in~\cite{pou97}, 
and extend the measurements to the 
SU(2)+Higgs case including now also the operator $G_\phi(x,y)$ 
(some properties of $G_\phi(x,y)$ have been  
previously studied in \cite{leip}).          
The discretization of the
operators in eqs.~\nr{GPhi}, \nr{GA}
is explained in Sec.~2. Some perturbative calculations 
are performed in Sec.~3.
The technical details of the simulations are given in Sec.~4 and
the numerical results are presented in Sec.~5.
In Sec.~6 we discuss
the physical implications of our results for the various topics sketched
above, and Sec.~7 comprises the conclusions.

\section{The lattice operators and the connection
to the static potential} 
 
The continuum Lagrangian of the 3d SU(2)+Higgs model is
\be
{\cal L} = 
\fr14 F^a_{ij}F^a_{ij}+(D_i\varphi)^\dagger D_i\varphi
+m_3^2\varphi^\dagger\varphi
+\lambda_3(\varphi^\dagger\varphi)^2.
\ee
On the lattice, one introduces  $\phi=(\tilde\varphi\;\varphi)$, 
where $\tilde\varphi=i\sigma_2\varphi^*$ and $\sigma_2$ is the Pauli 
matrix. 
After a rescaling of $\phi$, the lattice 
action is ($P_{ij}$ is the plaquette)
\ba
S[\phi,U]&=& \beta_G \sum_x \sum_{i<j}[1-\fr12 \tr P_{ij}] 
 - \beta_H \sum_x \sum_{i}
\fr12\tr\phi^\dagger(x)U_i(x)\phi(x+\hat i) \nonumber \\
 &+& \sum_x
(1-2\beta_R)\fr12\tr\phi^\dagger(x)\phi(x) + \beta_R\sum_x
 \bigl[ \fr12\tr\phi^\dagger(x)\phi(x)\bigr]^2.
\la{actlat}
\ea
In order to make contact with the calculations in \cite{ptw96}
we fix the ratio of the continuum scalar and gauge couplings to be
\beq
 x\equiv \frac{\lambda_3}{g_3^2} = \frac{\beta_R\,\beta_G}{\beta_H^2}=0.0239.
\eeq
At this parameter value the Higgs model exhibits a strong first-order 
phase transition upon variation of $\beta_H$, or
$y\equiv m_3^2/g_3^4$ in continuum notation. 
We pick the same two points
in parameter space as in \cite{ptw96}, 
namely $y=0.089$ and $y=-0.020$,
representing the confinement and Higgs phase, respectively.
Once $\beta_G$ is chosen, the appropriate values
of $\beta_H$ and $\beta_R$ are
determined by the ``lines of constant physics'' which 
govern the approach of the 
3d theory to the continuum limit \cite{fkrs95}. 
Results for the pure gauge theory may be obtained by 
considering only gluonic operators and simulating at $\beta_H=0$.

Consider then the gauge-invariant two-point function in 
eq.~\nr{GPhi}. The path $\Gamma$ between $x$ and $y$ could
be for instance a rectangular one as shown in Fig.~1(a). 
However, it has been demonstrated by simulations in the 
4d SU(2)+Higgs model that the coefficient $M$ of the exponential
decay of $G_\phi(x,y)$ is independent of $R$~\cite{ev86}. 
Hence, we restrict 
attention to $R=0$ in the following. 
Choosing $T$ to be in the 3-direction 
the simplest choice representing eq.~\nr{GPhi} is then
\beq \label{scallat}
G_{\phi}(T)= \left \langle \fr12\tr \left [\phi^{\dag}(x)
S(x,T)
\phi(y)\right] \right \rangle, 
\eeq
with 
\beq \label{sdef}
S(x,T)=\prod_{n=0}^{N-1} U_{3}(x+n\hat{3}), \qquad y=x+N\cdot \hat{3},
\eeq
and $T=N a$. In practice, we average this operator over the whole lattice 
in order to improve statistics.
One can also choose fundamental charge
operators other than the $\phi$'s at the ends of the Wilson line
(see Sec.~4). The expectation (to be tested below)
is that for large $T$, $G_\phi(T)$ should decay 
as $\exp{(-M\,T)}$ both in the Higgs and
in the screened confinement phase~\cite{bri83}.

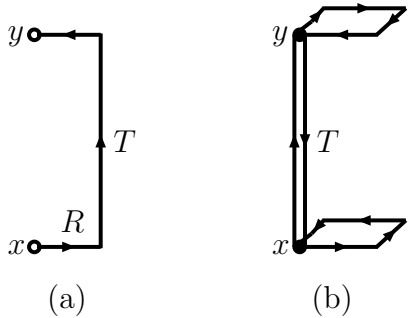
\begin{figure}[t] 

\vspace*{-0.5cm}

\begin{center}

\begin{picture}(200,120)(0,0)
 
\SetWidth{1.5}
\GCirc(20,20){2}{1}
\GCirc(20,100){2}{1}
\ArrowLine(21,20)(45,20)
\ArrowLine(45,19.5)(45,100.5)
\ArrowLine(45,100)(21,100)
 
\GCirc(120,20){2}{0}
\GCirc(120,100){2}{0}
\Line(124.5,25)(118,20)
\ArrowLine(129,30)(124.5,25)
\ArrowLine(160,30)(128.5,30)
\ArrowLine(149,20)(160,30)
\ArrowLine(122,20)(149.5,20)
\ArrowLine(122,100)(122,20)
\ArrowLine(149.5,100)(122,100)
\ArrowLine(160,110)(149,100)
\ArrowLine(128.5,110)(160,110)
\Line(118,100)(124.5,105)
\ArrowLine(124.5,105)(129,110)
\ArrowLine(118,20)(118,100)

\Text(10,20)[l]{$x$}
\Text(10,100)[l]{$y$}
\Text(110,20)[l]{$x$}
\Text(110,100)[l]{$y$}
\Text(50,60)[l]{$T$}
\Text(127,60)[l]{$T$}
\Text(30,30)[l]{$R$}
\Text(25,0)[l]{(a)}
\Text(125,0)[l]{(b)}
 
\end{picture}

\vspace*{-0.5cm}

\end{center}
\caption[]{\label{path}\it The operators measured on the lattice:
(a) The correlator $G_\phi(x,y)$ with a fundamental Wilson line 
(in practice we have $R=0$, and the correlator is denoted
by $G_\phi(T)$). (b) The correlator 
$G_{F}(T)$ with an adjoint Wilson line.}

\vspace*{0.5cm}

\end{figure}  

The connection to the static potential and to screening 
is provided by employing the standard picture of string breaking 
through matter pair production. Assuming that the expectation
value of the Wilson loop has a perimeter law behaviour 
for a large closed loop $C$, 
$W(R,T) \propto \exp(-M|C|)=\exp[-2M(R+T)]$, 
the static potential, defined as  
$\widetilde V(R)=-\lim_{T\to\infty}
[\ln W(R,T)]/T$ or
\be
V(R)=-\lim_{T\to\infty}
\ln\left[\frac{W(R,T)}{W(R,T-1)}\right], 
\ee
obeys
\beq \label{asymp}
V(R\rightarrow\infty)=2 M.
\eeq
Here $M$ is assumed to be the same mass parameter as 
measured from eq.~\nr{scallat}, when an optimal choice is made for the
operator at the ends of the Wilson line.
The physical interpretation of eq.~\nr{asymp}
is that at infinite separation of the static sources, the static potential
consists just of the energy of the hadronized system with each static
source binding a dynamical charge.

Consider then the case of a 
static source in the adjoint representation.
The Wilson line in the time 
direction now has 
to be taken in the adjoint representation, and
the gluon field binding to it is described by a spatial plaquette.
Hence we consider the gauge-invariant correlator 
shown in Fig.~1(b), 
\beq \label{wadj}
G_{F,ijkl}(T)=
\left\<
4 \tr ( P_{ij}(x) T^a) 
\Gamma^{ab}(T) 
\tr ( P^{\dag}_{kl}(y)T^b)\right\>, 
\la{GF}
\eeq 
where
\beq
\Gamma^{ab}(T)=2\tr \left (T^a S(x,T) T^b S^{\dag}(x,T) \right),
\eeq
and $S$ and $y$ are as in eq.~(\ref{sdef}).
This may be rewritten as
\ba
G_{F,ijkl}(T) & = & \left \langle 
2  \tr \left [ P_{ij}(x)S P^{\dag}_{kl}(y)S^{\dag} 
\right ] - \frac{2}{N} \tr P_{ij}(x) \tr P_{kl}(y) \right \rangle \nn 
& =  & \left \langle \tr \left [ P_{ij}(x)S \left ( P^{\dag}_{kl}(y)-
P_{kl}(y) \right ) S^{\dag}\right ]\right \rangle ,
\ea
where the last equality holds for SU(2) only. In our simulation,
we use $G_F$ in this last form, with the components $ij=kl=12$;
we omit the spatial indices of $G_F(T)\equiv G_{F,1212}(T)$ in the following.
The plaquettes are replaced by the 
sum over all four spatial plaquettes of the same orientation sharing
the end points of the Wilson lines, 
i.e.~$P(x)\equiv P_{12}(x)+P_{12}(x-\hat{1})
+P_{12}(x-\hat{2})+P_{12}(x-\hat{1}-\hat{2})$.
To improve on the statistics, we again 
average over the whole lattice. 
Adjoint charge operators other than $P_{ij}$
can be considered as well (see Sec.~4). 

The relation to the static potential may be
taken over from the fundamental case. 
It is known that the static potential in this case also shows a
linear rise due to flux tube formation between the static charges.
As the flux tube now is in the adjoint representation it couples to the
gauge fields and thus is expected to 
break even in the case of pure gauge theory.
The corresponding final state consists of the static adjoint source binding
a dynamical gluon, a system that has been termed ``glue-lump" in the
literature \cite{mic85,pou97}. 

\section{Perturbation theory}

Although the purpose of this paper is a non-perturbative
measurement of the correlators in eqs.~\nr{GPhi}, \nr{GA}, 
let us in this section study these quantities perturbatively. 
The main motivation is to see how the mass parameters measured 
depend on the lattice spacing $a$. We will also make some 
other computations in the Higgs and confinement phases. 

\subsection{Scaling with the lattice spacing\label{div}}

\begin{figure}[tbh] 

\vspace*{-0.5cm}

\begin{center}

\begin{picture}(150,120)(0,0)
 
\SetWidth{1.5}
\ArrowLine(22,20)(22,100)
\PhotonArc(22,60)(20,-90,90){1.5}{8}

\ArrowLine(122,20)(122,100)
\ArrowLine(118,100)(118,20)
\PhotonArc(122,60)(20,-90,90){1.5}{8}

\Text(14,0)[l]{(a)}
\Text(114,0)[l]{(b)}
 
\end{picture}

\vspace*{-0.5cm}

\end{center}
\caption[]{\label{wpath}\it 
The Wilson line self-energy
in the fundamental (a) and adjoint (b) representation.}

\vspace*{0.5cm}

\end{figure}
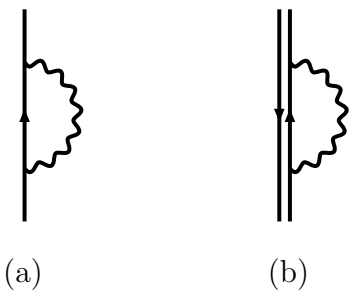

The mass parameters measured from eqs.~\nr{GPhi}, \nr{GA}
turn out to contain a divergent part $\propto g_3^2 \ln(a)$. 
There is no counterterm in which to absorb this divergence
so that, in fact, there is no meaningful continuum limit. 
Nevertheless, measurements with a finite lattice spacing 
can be useful for 
a number of applications, as we will see. Due
to the fact that the gauge coupling is dimensionful in 3d, the 
divergent part can be determined with a 1-loop 
computation. Moreover, it can be seen that the only 
such contribution comes from the Wilson line,  
depicted in Fig.~\ref{wpath}. Following \cite{ay}, 
let us thus consider this object.

Letting the Wilson line be an infinitely 
long straight line in the 
$z$-direction, expanding the 
path-ordered
exponential to second order
in the fields and taking the derivative with respect to $z$, one finds   
for the coefficient $M$ in the exponential fall-off, 
\ba
M & = &  g_3^2 \lim_{z\to\infty} \frac{d}{dz}
\left[\int_0^z\! dy\int_0^y\! dx
\langle A^a_z(x,{\bf 0}) A^b_z(y,{\bf 0})\rangle
\right] (T^aT^b)_{\alpha\beta} \nn
& = & g_3^2 \int_0^\infty dx 
\langle A^a_z(x,{\bf 0}) A^b_z(0,{\bf 0})\rangle (T^aT^b)_{\alpha\beta}\nn
& = & 
g_3^2 C_2\fr12 \int \frac{d^2 p_\perp}{(2\pi)^2}
\frac{1}{p_\perp^2+m_W^2}. \la{E}
\ea
Here $\alpha=\beta$ are isospin indices not summed over, 
$C_2=N$ for the adjoint representation, 
and $C_2=(N^2-1)/(2N)$ for the fundamental representation. 
The gauge field propagator was taken in a general
gauge in the Higgs phase of the SU(2)+Higgs model with 
the mass $m_W$. The result in eq.~\nr{E} applies also on the lattice
when the integration range is restricted to $(-\pi/a,\pi/a)$
and the momentum in the propagator is replaced with 
$p_\perp^2\to \tilde p_\perp^2 = \sum_{i=1}^2 \tilde p_i^2$, 
where $p_i = ({2}/{a})\sin ({a p_i}/{2})$. 
The intergal in eq.~\nr{E} can be computed and one gets
\ba
\int_{-\pi/a}^{\pi/a}\!\!\frac{d^2p}{(2\pi)^2}
\frac{1}{\tilde p^2+m^2} & = & \frac{1}{4\pi}
\ln\left(\frac{32}{m^2 a^2}\right)+{\cal O}(a) \quad\quad ({\rm lattice}),
\la{intlat} \\
\int\!\!\frac{d^{2-2\epsilon}p}{(2\pi)^{2-2\epsilon}}
\frac{1}{p^2+m^2} & = & \frac{\mu^{-2\epsilon}}{4\pi}
\left(\frac{1}{\epsilon}+
\ln\frac{\overline\mu^2}{m^2}\right)
\quad\quad\quad (\mbox{\rm $\overline{\mbox{\rm MS}}$ in continuum}),
\ea
which provides a relation between the two schemes. 
Eqs.~\nr{E}, \nr{intlat} 
tell how the mass parameters measured on
the lattice depend on the lattice spacing.

In the following, we will want to consider quantities which
do have a continuum limit. From eqs.~\nr{E}, \nr{intlat}, it 
can be seen that this is obtained by subtracting from $M$
a divergent part: 
\be
{M'}\equiv {M}-{g_3^2}\frac{C_2}{8\pi}
\ln\left(\frac{32}{g_3^4a^2}\right)=
{M}-{g_3^2}\frac{C_2}{4\pi}
\ln\left(\frac{2\sqrt{2}\beta_G}{N}\right),
\la{subtraction}
\ee
where $\beta_G=2 N/(g_3^2 a)$ for SU(N).
Here the mass scale needed to define the 
logarithm in eq.~\nr{intlat}
was chosen to be $g_3^2$.
The quantity $M'$ is a continuum quantity in the sense that 
in the $\overline{\mbox{\rm MS}}$ scheme 
with the scale 
$\overline{\mu}=g_3^2$, the exponential fall-off is
determined by 
\be
M_{\overline{\mbox{\rm\tiny MS}}}=
M'+g_3^2\frac{C_2}{8\pi}\frac{1}{\epsilon}.
\ee

\begin{figure}[tbh] 

\vspace*{-0.5cm}

\begin{center}

\begin{picture}(300,120)(0,0)
 
\SetWidth{1.5}
 
\GCirc(25,20){2}{1}
\GCirc(25,100){2}{1}
\DashLine(25,21)(25,99){5}
 
\GCirc(100,20){2}{1}
\GCirc(100,100){2}{0}
\GCirc(125,20){2}{1}
\GCirc(125,100){2}{0}
\GCirc(150,20){2}{1}
\GCirc(150,100){2}{0}
\GCirc(175,20){2}{1}
\GCirc(175,100){2}{0}
\DashLine(100,21)(100,50){5}
\DashLine(125,21)(125,50){5}
\DashLine(150,21)(150,50){5}
\DashLine(175,21)(175,50){5}
\Line(96,46)(104,54)
\Line(96,54)(104,46)
\DashCArc(125,60)(10,0,360){5}
\PhotonArc(150,60)(10,0,360){1.5}{8}
\CArc(175,60)(10,0,360)
\CArc(175,60)(7,0,360)
 
\GCirc(240,20){2}{0}
\GCirc(240,100){2}{0}
\ArrowLine(240,20)(240,100)
\PhotonArc(240,60)(20,-90,90){1.5}{8}
 
\Text(17,0)[l]{(a)}
\Text(130,0)[l]{(b)}
\Text(235,0)[l]{(c)}
 
\end{picture}

\vspace*{-0.5cm}

\end{center}
\caption[]{\label{path2}\it 
The graphs contributing to the correlator $G_\phi(T)$
in the Higgs phase at 1-loop order. A dashed line is 
a Higgs field, a wiggly line a gauge field, a double
line a ghost field and the cross denotes a counterterm. A filled 
circle is the shifted field $\hat\varphi$, whereas an open
circle is the quantum field $\varphi'$.}

\vspace*{0.5cm}

\end{figure}
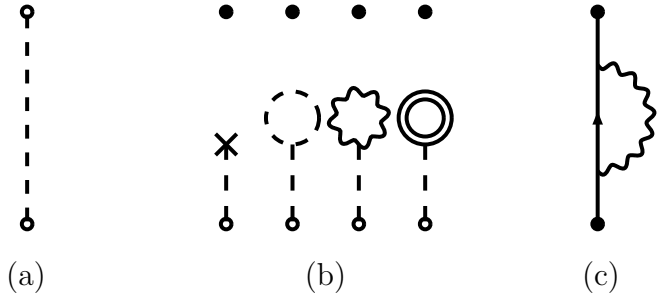  

\subsection{Higgs phase\label{Hphase}}

Let us then consider the correlator
$G_\phi(T)$ in the Higgs phase of the SU(2)+Higgs model
in some more detail.
The Higgs field is thus written as $\varphi=\hat\varphi+\varphi'$, 
$\hat\varphi\neq 0$. At leading order, one gets just the 
constant $\hat\varphi^\dagger\hat\varphi$. The graphs 
contributing at the next order are shown in Fig.~3. 
The other graphs are straightforward, so let us consider
the Wilson line contribution, (3.c). For arbitrary $T$, we get
\ba
\left. G_\phi(T)\right|_{\rm (3.c)}& = & \fr32\biggl[
-f(\xi^{1/2}m_W,T)
+f(\xi^{1/2}m_W,0)
+f(m_W,T)-f(m_W,0) \nn
& &   
-2 \int_{-\pi/a}^{\pi/a}\!\!\frac{d^3p}{(2\pi)^3}
\frac{m_W^2}{\tilde p^2+m_W^2}
\frac{\sin^2(\frac{p_z T}{2})}{\tilde p_z^2}
\biggr]. \la{dGPhi}
\ea
Here $\xi$ is the gauge fixing parameter of an $R_\xi$ gauge and
\be
f(m,|z|)=
\int_{-\pi/a}^{\pi/a}\!\!\frac{d^3p}{(2\pi)^3}
\frac{e^{ip\cdot z}}{\tilde p^2+m^2},
\ee
with the limiting values ($\Sigma = 3.175911535625$)
\be
f(m,0)=\frac{\Sigma}{4\pi a}-\frac{m}{4\pi}+ O(am^2), \quad
f(m,z)\stackrel{|z|\gg 1/m}{\longrightarrow} \frac{e^{-m|z|}}{4\pi |z|}.
\ee
The gauge parameter dependent parts in eq.~\nr{dGPhi} cancel those form 
the other graphs, so that the result is explicitly gauge
independent. In the limit $T\to\infty$, 
eq.~\nr{dGPhi} is dominated by a linear term, 
$\left. G_\phi(T)\right|_{\rm (3.c)} \to 
\hat\varphi^\dagger\hat\varphi (-MT)$, 
where $M$ is from eqs.~\nr{E}, \nr{intlat} with $C_2=3/4$. 

Including also the other
graphs, the complete 1-loop answer for large $T$ is
\ba
G_\phi(T) &=& \hat\varphi^\dagger\hat\varphi
\left[1-T \frac{3 g_3^2}{16\pi} \ln\frac{4\sqrt{2}}{m_W a}\right] \nn
&+& \frac{\Sigma}{4\pi a}\left(1+\fr34 \frac{g_3^2}{\lambda_3}\right)
-f(m_H,0)-\fr34\frac{g_3^2}{\lambda_3}f(m_W,0) \nn
&+& \frac{1}{8\pi} m_H + \frac{3}{4\pi} m_W + 
\frac{e^{-m_H T}}{8\pi T}+\frac{3}{4\pi}\frac{e^{-m_W T}}{m_WT^2}
+O(a)+O\left(\frac{1}{m T}\right), \la{pertGPhi}
\ea
where $m_W, m_H$ are the perturbative $W$ and Higgs masses, 
respectively. The $1/a$-term on the 2nd row cancels the 
divergences from $f(m_H,0)$, $f(m_W,0)$, so that the 
sum is finite.  
At $T=0$, on the other hand, 
the result is the first two rows plus 
a divergent term $\Sigma/(2\pi a)$.

In the Higgs phase for $G_\phi(T)$, 
one can thus see explicitly how a pure exponential
behaviour arises, after the exponentiation
into $\hat\varphi^\dagger\hat\varphi\exp(-MT)$ of 
the first row in eq.~\nr{pertGPhi}
(analytic arguments for the exponentiation can be    
obtained with the cumulant expansion, see, e.g., \cite{cum}).    
This is in contrast to the typical perturbative
terms on the 3rd row containing pre-exponential corrections.
It will be seen that the pure exponential is
indeed what is observed on the lattice. 

For the vector correlator in the Higgs phase, the 
corresponding computation is much more tedious. 
At the continuum limit one may write
\be
P_{ij}(x) \approx e^{ia^2g_3T^aF^a_{ij}(x)}
\ee
in eq.~\nr{GF}, 
and the leading term in the $\overline{\mbox{\rm MS}}$ scheme is 
\be
G_F(T) \propto 
\langle F^a_{12}(0)F^a_{12}(T)\rangle \approx 
(N^2-1)\frac{m_W}{2\pi T^2}e^{-m_W T}. \la{loGF}
\ee
Taking into accout the next order,
one gets a contribution $\exp(-M^{\rm adj}T)$ from the 
Wilson line multiplying eq.~\nr{loGF}, where,
according to eqs.~\nr{E}, \nr{intlat}, 
\be
 M^{\rm adj} = \frac{g_3^2}{2\pi}\ln\frac{4\sqrt{2}}{a m_W}.
\ee
However, there 
are many other contributions as well (in pure 4d SU(N) 
in the continuum, 
the graphs have recently been  
computed in~\cite{jamin}). 
We will not evaluate these
graphs explicitly. It will be seen that on 
the lattice the behaviour of $G_F(T)$ is again favoured to be purely 
exponential rather than with a prefactor as in eq.~\nr{loGF}, 
and that the exponent is $\sim m_W+M^{\rm adj}$.

\subsection{Confinement phase}

Consider then the confinement phase of the SU(2)+Higgs model,  
or the pure SU(2) theory. 
From eqs.~\nr{E}, \nr{intlat}
it is immediately seen that a perturbative 
computation of the correlators
is not possible: the tree-level mass 
parameter $m_W$  vanishes so that the IR-regime 
makes $M$ divergent. One can only extract the coefficient
of the logarithm which determines how the mass $M$ 
scales when the lattice spacing $a$ is varied. 

There is, however, the following meaningful 
computation one can make~\cite{ay}. 
(These considerations are in complete analogy with those
for the heavy -- light quark system in 4d, 
see~\cite{ei88}.) To be specific, consider
the 3d SU(2)+adjoint Higgs theory, 
\be
L= \fr14 F^a_{ij}F^a_{ij}+\fr12(D_iA_0)^a(D_iA_0)^a+
\fr12 (m_D^\rmi{LO})^2 A_0^aA_0^a + 
\fr14 \lambda_A(A_0^aA_0^a)^2, \la{adact}
\ee
where the mass
parameter $m_D^\rmi{LO}$ is large. Suppose one wants
to compute the mass $M_{HL}$ of the ``heavy--light'' bound state 
$h_i=\epsilon_{ijk} A_0^a F^a_{jk}$. A perturbative computation meets
immediately with IR-problems~\cite{ay,rebhan}. In addition, 
lattice simulations in the full theory are difficult 
since the requirement $aM_{HL}\ll 1$ is very stringent, 
due to the heavy constituent mass $m_D^\rmi{LO}$
(see, however, \cite{adjoint2}). On the
other hand, perturbation theory does work for integrating
out $A_0$, since there are no IR-problems due to the
large mass. Therefore, one can integrate $A_0$ out
analytically, replacing $\<h_i(0) h_i(T)\>$  by precisely the 
correlator $G_F(T)$ in eq.~\nr{GA} in the pure 3d SU(2) theory, 
times an overall factor. 
The only subtlety is that the heavy mass 
$M_H\sim m_D^\rmi{LO}$ appearing in the overall factor
gets modified in the integration procedure. As a result, 
the mass signal $M$ measured from $G_F(T)$ represents
the difference
\be
M = M_{HL}-M_H, \la{mHLmH}
\ee
where $M_H$ is determined by the computation in~\cite{ay} to be 
(for $\lambda_A\ll g_3^2$) 
\be
M_H = m_D^\rmi{LO} + \frac{Ng_3^2}{8\pi}
\left[\ln\frac{(a m_D^\rmi{LO})^2}{8}-1\right]. \la{MH}
\ee
Then a lattice measurement of $M$  
from $G_F(T)$
in the simpler pure SU(2) gauge
theory allows a non-perturbative determination of $M_{HL}$, 
using eq.~\nr{mHLmH}.
These considerations are directly relevant for the numerical determination
of the finite temperature Debye mass $m_D$ in SU(N) QCD;
we will return to this subject in Sec.~6.1.  

A computation similar to that for the adjoint Wilson line
described above 
could be carried out for the fundamental Wilson line
in the SU(2)+Higgs model, as  
well. This would allow a determination of the mass
of the bound state $\varphi^\dagger\chi$ in a 3d theory
with two SU(2) Higgs doublets $\varphi$, $\chi$, 
of which $\chi$ is heavy. In the limit that $\phi$ and $\chi$
interact only through gauge interactions, the
computation is quite analogous
to the one above, with the change $N \to (N^2-1)/(2N)$  
relevant for the fundamental representation.

\section{Simulations and analysis} 

The algorithm used to perform the Monte Carlo simulation using the
action in eq.~(\ref{actlat}) is the same as in~\cite{ptw97,ptw96}. 
The gauge variables are updated
by a combination of heatbath and over-relaxation steps
according to \cite{fab,ken}, while the scalar degrees of freedom
are updated combining heatbath and reflection steps
as described in \cite{bunk}.
The ratio of the different updating steps is suitably tuned
such as to minimize autocorrelations. 
In our simulations we typically gathered between 5000 and 10000
measurements taken after such combinations of updating sweeps.

In order to investigate the scaling properties of the mass parameters 
and to extra\-polate to the continuum we have  
simulated with $\beta_G=9,12,16$ ($\beta_G=4/g_3^2a$).
For each value of $\beta_G$ we have
checked that the extracted mass 
parameters are free from finite size effects, using
lattices ranging from $14^2\cdot 20$ to $54^3$.

\subsection{Smearing and matrix correlators}

As mentioned in Sec.~2, the 
choice of operators at the ends of the Wilson line
in eqs.~\nr{scallat}, \nr{GF} is not unique. One could
choose any other operators in the same representation 
as well, and the task is to find those operators
which give the smallest mass parameters
(i.e., which couple to the ``ground state''). 
A systematic way
of doing this is to take a set of trial operators and to
measure the whole correlation matrix, which can then be diagonalized.

To form 
the basis of operators,
one possibility is to take ``smeared'' or ``blocked'' fields. 
It has been demonstrated that 
using smeared variables instead of
the original ones greatly improves the projection properties
of gauge-invariant operators employed in calculations of the 
mass spectrum~\cite{ptw96}. 
Similar findings have been reported from calculations
of the adjoint Wilson line correlator in eq.~(\ref{wadj}) 
\cite{mic85,pou97}. 

To be specific, we construct link variables 
of blocking level $n$ according to \cite{tep87}
\bea \label{lbl}
U_{i}^{(n)}(x)&=&\fr13 
\left\{U_i^{(n-1)}(x)U_{i}^{(n-1)}(x+\hat{i}) \right.\nn
&&+\left.\sum_{j=\pm 1,j\neq i}^{\pm 2}
U^{(n-1)}_j(x)U_i^{(n-1)}(x+\hat{j})
U_i^{(n-1)}(x+\hat{i}+\hat{j})U_j^{(n-1)\dagger}(x+2\hat{i})\right \},
\eea
and composite scalar variables of blocking level $n$ 
as in \cite{ptw96},
\beq \label{sbl}
\phi^{(n)}(x)=\fr15 \left\{\phi^{(n-1)}(x)+\sum_{i=1}^{2}\left[
U_i^{(n-1)}(x)\phi^{(n-1)}(x+\hat{i})+U_i^{(n-1)\dagger}(x-\hat{i})
\phi^{(n-1)}(x-\hat{i})\right]\right\},
\eeq
where $i=1,2$,
i.e., smearing is performed in the spatial plane. The Wilson
lines connecting the fields are in the time direction and remain
unsmeared. 
Possible improvement techniques for the time-like links have been 
discussed, e.g., in~\cite{mic85,pou97}.

The diagonalization of the correlation matrix is performed using 
a variational method 
(see, e.g., \cite{mic85,ptw97,ptw96}). 
For this purpose we iterate our spatial smearing procedure four times
and measure the $5\times 5$ correlation matrices
\bea
G^{nm}_{\phi}(T)&=& \left \langle \fr1V \sum_x \fr12\tr 
\left [\phi^{(n)\dag}(x)
S
\phi^{(m)}(y)\right ] \right \rangle,\\
G_F^{nm}(T)&=&\left\langle \fr1V \sum_x
\tr \left [P^{(n)}(x)S \left( P^{(m)\dag}(y)-
P^{(m)}(y)\right) S^{\dag}\right ] \right\rangle,
\eea
where the plaquettes in $P^{(n)}$ have been constructed out of 
smeared links at blocking level $n=1,\dots,5$.
For a given set of smeared scalar fields, we find the linear combination
$\Phi_1=\sum a_{1k} \phi^{(k)}$
that maximizes $G_{\phi}(a)$, thus 
isolating the linear combination giving the lightest
mass parameter. 
The first ``excitation'' may be found by repeating this step restricted
to the subspace $\{\phi^{(n)'}\}$ which is orthogonal to the 
ground state. 
Hence we can, in principle, 
obtain five eigenstates of the matrix correlator 
given by
\beq
\Phi_i =  \sum_{k=1}^5 a_{ik}\phi^{(k)}.
\eeq
Our mass estimates are then obtained from the correlation functions
${G}_{\Phi_i}(T)$ 
calculated in the diagonalized basis.
Since the basis gets smaller for 
higher excitations, the reliability of the mass estimates 
rapidly deteriorates for higher states. Of course, one could improve on this
by extending the basis of operators.
However, even a small operator basis has been 
demonstrated to work quite well in the
case of the lowest gauge-invariant
eigenstates \cite{ptw96}. 
In the following we shall present results for the ground states and,
where the reliability seems to be reasonable,  
the first excitation. 

Note that, in contrast to calculations of correlation functions of 
gauge-invariant operators, the eigenvectors
$\Phi_i$ themselves are not gauge-invariant and hence do not represent
eigenvectors of the lattice Hamiltonian describing physical states.
Nevertheless, the $\Phi_i$ are gauge-covariant and  
the variational procedure will work to minimize
the exponents of the fall-off. 
The same considerations hold for the field strength correlator
(see also~\cite{mic85}).

\subsection{Fitting functions and error analysis}

An important question concerns the fitting functions to be chosen
in order to obtain mass estimates from the two-point functions. 
Unlike in the case of typical spectrum measurements where 
time slice correlators projecting on zero transverse momentum are considered, 
it is not a priori clear which asymptotic form the non-local
operators take for large $T$. In particular, perturbation theory 
suggests that there are terms with exponential decay modified by
power law corrections, see eqs.~(\ref{pertGPhi}),(\ref{loGF}).
On the other hand, according to \cite{bri83}, one 
expects a pure exponential decay. 
In order to clarify this question we consider three fitting functions,
$\exp(-MT)/T^{\alpha}$ with $\alpha$ as an open parameter, 
and the special cases  
$\exp(-MT)/T$ and $\exp(-MT)$.
As we shall see, a pure exponential is the asymptotic form preferred by the
diagonalized data
(for the undiagonalized data the eventual conclusion is the same, 
but to achieve it requires a more tedious analysis).

Our final mass estimates were therefore obtained by performing correlated
fits of the form $\sim \exp(-MT)$ over some interval $[T_1,T_2]$
to the diagonalized correlators ${G}_{\Phi_i,F_i}(T)$.
We have checked 
our results for stability under variations of the fitting interval,
and also for compatibility with the results of uncorrelated fits. 
In those 
cases where different fitting procedures
gave results
that were not compatible within errors 
we quote the discrepancy as a systematic error.

\section{Numerical results}

In this section, we present our main numerical results. 
In Sec. \ref{asy} we demonstrate 
the necessity of smearing and diagonalization
in order to get a satisfactory signal with reasonable computer
resources, and discuss the asymptotic form of the correlators.
The mass estimates obtained for 
$\beta_G=9,12,16$ and for various lattice volumes are contained in
Sec.~\ref{mass} and the extrapolation to the continumm
limit is discussed in Sec.~\ref{cont}. 

\subsection{\label{asy} Asymptotic form of the correlators}

\begin{figure}[tbp]

\vspace*{-1.2cm}

\centerline{\hspace{-2mm}
\epsfxsize=9.5cm\epsfbox{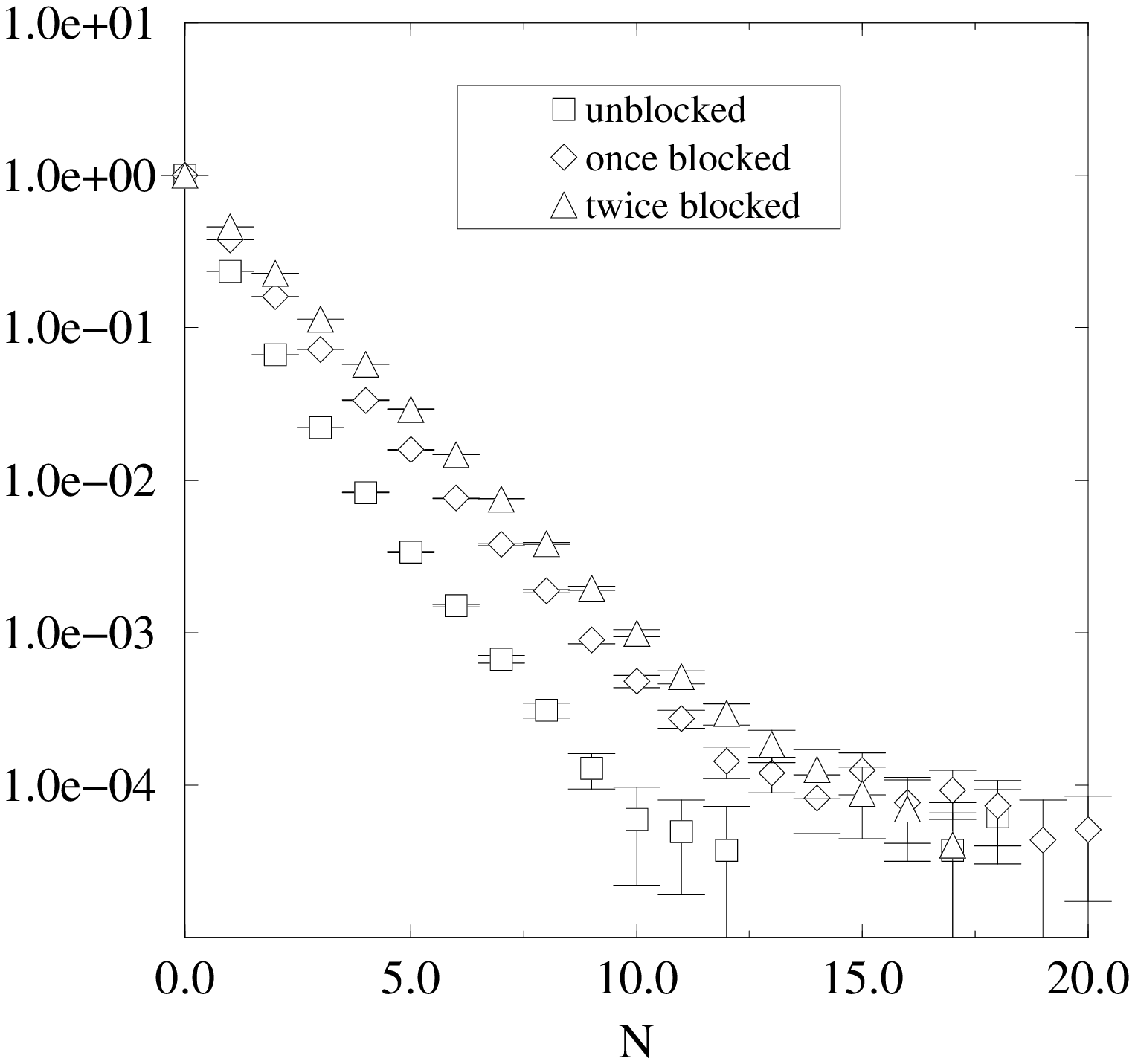}
\hspace{-1.5cm}
\epsfxsize=9.5cm\epsfbox{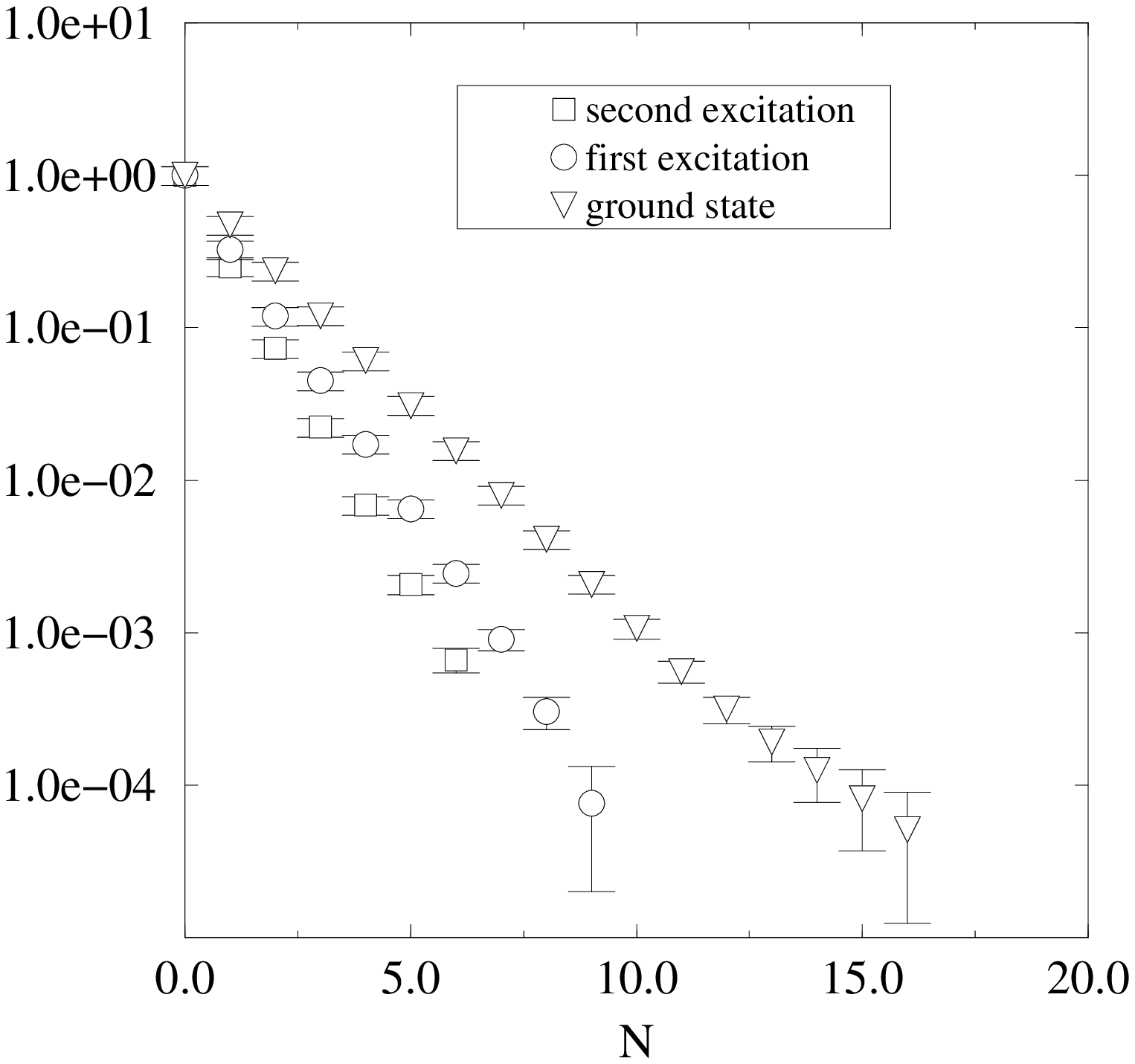}}

\vspace*{-4.5cm}

\caption[]{\label{corgf}
{\it Left: The correlator $G_F(T)$ for three different blocking levels
     in the confinement phase.
     Right: The diagonalized correlators ${G}_{F_i}(T)$ for the three lowest 
         states.  Here $\beta_G=9$, volume $=42^3$, $T=Na$.}}
\end{figure}

An example of the effects of the smearing and diagonalisation procedures 
is shown in Fig.~\ref{corgf}. The field strength correlator  
is depicted in the confinement phase at $\beta_G=9$ on a $42^3$ lattice.
This corresponds to the largest volume in physical units that we have 
considered. On the left the unblocked correlator is shown together with
the once and twice blocked versions, all normalized to one at zero distance.
It is immediately apparent that the unblocked correlator exhibits 
a faster decay at small distances 
than the blocked ones, and hence it is more difficult 
to extract the asymptotic mass value, 
although the data seem to be quite good.
Furthermore, it turns out that all three correlators 
have a slight curvature. 
It requires a careful analysis of the stability of $M$ with respect 
to different fitting ranges in order to decide that
$\exp(-MT)$ gives a better fit than, 
for example, $\exp(-MT)/T$.

%

\begin{table}[tb]
\begin{center}
\begin{tabular}{||l||r@{.}l|r@{.}l|r@{.}l||r@{.}l|r@{.}l||r@{.}l|r@{.}l||}
\hline
\hline
  & \multicolumn{6}{c||}{$\exp(-MT)/T^{\alpha}$}
  & \multicolumn{4}{c||}{$\exp(-MT)/T$}
  & \multicolumn{4}{c||}{$\exp(-MT)$}  \\
\hline
\hline
$T_1,T_2$ & \multicolumn{2}{c|}{$aM$}
              & \multicolumn{2}{c|}{$\alpha$}
              & \multicolumn{2}{c||}{$\chi^2$/dof}
              & \multicolumn{2}{c|}{$aM$}
              & \multicolumn{2}{c||}{$\chi^2$/dof} 
              & \multicolumn{2}{c|}{$aM$}
              & \multicolumn{2}{c||}{$\chi^2$/dof}\\
\hline
7, 12 & 0&71(11) & -0&30(84)& 0&63 & 0&5366(96) & 1&15 & 0&6700(93) & 0&50 \\
6, 12 & 0&655(55)& 0&17(35) & 0&58 & 0&5217(50) & 1&55 & 0&6758(46) & 0&51 \\
5, 12 & 0&651(24)& 0&16(14) & 0&63 & \multicolumn{2}{c|}{\mbox{[huge error]}}&
\multicolumn{2}{c||}{\mbox{--}} & 0&6795(29) & 0&61 \\
\hline
\hline
\end{tabular}
\caption{\label{fits} {\it 
Comparison of fits to ${G}_{F_1}(T)$ in the confinement phase 
with different functions over various fitting ranges at $\beta_G=9$, 
$L^2\cdot L_T=42^3$.}}
\end{center}
\end{table}

The situation can be significantly improved upon by performing the 
diagonalization outlined
in the last section. The diagonalized correlation functions corresponding 
to the three lowest states are shown on the RHS in Fig.~\ref{corgf}.
In Table \ref{fits} we compare the results of fitting different functions to 
the diagonalized 
correlator with the lowest mass, ${G}_{F_1}(T)$. Acceptable fits 
using $\exp(-MT)/T$ may be obtained on the late timeslices. 
However, if one extends the fitting range to earlier time slices the
quality of the fit rapidly deteriorates, giving large $\chi^2$/dof and/or
large errors. 
On the other hand, we obtain good fits with consistent mass values
using a pure exponential $\exp(-MT)$ over a larger range of fitting intervals.
Finally, three-parameter fits employing
the functional form $\exp(-MT)/T^{\alpha}$ yield values for $\alpha$ that
are consistent with zero as well as mass values consistent with those of 
pure exponential fits. The same behaviour is found for the scalar field
correlator and in the Higgs phase as well. The fits for these cases
are shown in Table~\ref{alphafits}.

\begin{table}[tb]
\begin{center}
\begin{tabular}{||l||r@{.}l|r@{.}l|r@{.}l||r@{.}l|r@{.}l||r@{.}l||}
\hline
\hline
  & \multicolumn{6}{c||}{$\exp(-MT)/T^{\alpha}$}
  & \multicolumn{4}{c||}{$\exp(-MT)$}  \\
\hline
\hline
              & \multicolumn{2}{c|}{$aM$}
              & \multicolumn{2}{c|}{$\alpha$}
              & \multicolumn{2}{c||}{$\chi^2$/dof}
              & \multicolumn{2}{c|}{$aM$}
              & \multicolumn{2}{c||}{$\chi^2$/dof}\\
\hline
${G}_{F_1}$, conf. &   
0&655(55)& 0&17(35)  & 0&58 & 0&6758(46) & 0&51 \\  
${G}_{\Phi_1}$, conf. &
0&3305(53) & -0&060(40) & 0&53 & 0&3228(13) & 0&72 \\
${G}_{F_1}$, Higgs &
1&19(15)   & -0&4(5) & 0&34 & 1&09(2) & 0&38 \\
${G}_{\Phi_1}$, Higgs &
0&05424(8) & 0&0005(6) & \multicolumn{2}{c||}{\mbox{--}} 
& 0&05428(4) & \multicolumn{2}{c||}{\mbox{--}} \\
\hline
\hline
\end{tabular}
\caption[]{\label{alphafits} {\it
Comparison of fitting functions. In each case we consider the same fitting
range for both functions. 
The data are for $\beta_G=9$ from the largest lattices considered in each case.
The last line corresponds to uncorrelated fits and
no $\chi^2/dof$ is quoted.}}
\end{center}
\end{table}

Thus, although there are a few cases where, e.g., a $1/T$ modification of the 
exponential decay cannot be strictly ruled out
based on a finite fitting interval and the value of $\chi^2$/dof alone,
c.f.~Table \ref{fits},
we have strong numerical evidence that  
both of the correlators $G_\Phi(T)$ and $G_F(T)$   
decay with a pure exponential $\exp(-MT)$ for large $T$, in both phases. All
the following mass estimates have been obtained by fitting 
a pure exponential to the diagonalized 
correlation functions ${G}_{\Phi_i,F_i}(T)$.

\subsection{\label{mass} Results for the masses}

The numerical results for the mass of the
``ground state'' in the Higgs phase are summarized in lattice units  
in the first block in Table \ref{results_h}. 
In \cite{ptw96} it was found that the physical
Higgs and W boson masses in the Higgs phase 
are free from finite volume effects
already for rather small lattices, and our choice of lattice sizes
is motivated by these findings.
As a safeguard, we have performed an explicit check for finite size effects
at $\beta_G=9$, and Table \ref{results_h} shows them to be absent.
The spatial
volumes at $\beta_G=12,16$ are larger than merely the scaled up 
versions of the smallest lattice 
for $\beta_G=9$, so we are confident to have reached the infinite
volume limit in those cases as well.
Note the very weak exponential decay of the scalar correlator. 

\begin{table}[ht]
\renewcommand{\baselinestretch}{1.}
\begin{center}
\begin{tabular}{||l|r|r@{.}l|l||r@{.}l|r@{.}l|r@{.}l|r@{.}l||}
\hline
\hline
\mbox{}
& $\beta_G$
& \multicolumn{2}{c|}{$\beta_H$}
& $L^2\cdot L_T$
& \multicolumn{2}{c|}{$aM_{\Phi_1}$}
& \multicolumn{2}{c|}{$aM_{\Phi_2}$} 
& \multicolumn{2}{c|}{$aM_{F_1}$}
& \multicolumn{2}{c||}{$aM_{F_2}$} \\
\hline
\hline
%
%
\mbox{Higgs} & 16 & 0&3396 & $26^2\cdot 36$ & 0&04086(4) &
\multicolumn{2}{c|}{\mbox{--}} & 0&653(6) & 
\multicolumn{2}{c||}{\mbox{--}}  \\
\cline{2-13}
\mbox{phase} & 12 & 0&3418 & $24^2\cdot 42$ & 0&04769(3) &
\multicolumn{2}{c|}{\mbox{--}} & 0&836(5) &
\multicolumn{2}{c||}{\mbox{--}} \\
\cline{2-13}
\mbox{} & 9 & 0&3450 & $20^2\cdot 36$ & 0&05428(4) &
\multicolumn{2}{c|}{\mbox{--}} & 1&09(2) &
\multicolumn{2}{c||}{\mbox{--}} \\
\mbox{} &   &\mcemptyr & $14^2\cdot 36$ & 0&05426(5)  &
\multicolumn{2}{c|}{\mbox{--}}  & 1&09(2) &
\multicolumn{2}{c||}{\mbox{--}} \\
\hline
\hline
%
%
\mbox{conf.} & 
16 & 0&3392 & $54^3$ & 0&187(1)(1)& 0&327(14) & 0&397(2)(5)& 0&557(7)\\
\mbox{phase} & 
& \mcemptyr& $42^3$ & 0&187(2) & 0&359(14) & 0&395(5) & 0&559(4)\\
\cline{2-13}
\mbox{} & 12 & 0&3411   & $42^3$ & 0&2470(7)& 0&437(7)(12) & 0&516(2) & 
     0&729(9)(15) \\
\mbox{} & & \mcemptyr & $30^2\cdot 42$ & 0&247(1)& 0&448(7) & 0&514(6) & 
     0&730(8) \\
\cline{2-13}
\mbox{} & 
  9 & 0&3438  & $42^3$ & 0&323(2) & 0&565(4) & 0&677(2) & 0&964(4) \\
\mbox{} &
& \mcemptyr & $24^2\cdot 36$ & 0&321(2) & 0&557(5) & 0&678(2) &
 0&965(6)(5)\\
\hline
\hline
%
%
\mbox{pure} & 
16 & \multicolumn{2}{c|}{\mbox{--}} & 
 $54^3$ & \multicolumn{2}{c|}{\mbox{--}} & 
\multicolumn{2}{c|}{\mbox{--}} &
0&397(3) & 0&566(5) \\
 \mbox{SU(2)} &  & \multicolumn{2}{c|}{\mbox{--}} & 
 $42^3$ & \multicolumn{2}{c|}{\mbox{--}} &
\multicolumn{2}{c|}{\mbox{--}} &
0&401(4) & 0&566(6)  \\
\cline{2-13}
\mbox{} & 
12 & \multicolumn{2}{c|}{\mbox{--}} & 
 $30^2\cdot 42$ & \multicolumn{2}{c|}{\mbox{--}} &
\multicolumn{2}{c|}{\mbox{--}} &
0&517(7) & 0&744(6) \\
\cline{2-13}
\mbox{} &
 9 & \multicolumn{2}{c|}{\mbox{--}} & 
 $30^3$ & \multicolumn{2}{c|}{\mbox{--}} & 
\multicolumn{2}{c|}{\mbox{--}} & 0&685(3)(7) & 0&960(10) \\
\hline
\hline
\end{tabular}
\caption{\it The coefficients of the exponential decay 
  in lattice units.
  The first error is statistical and the
  second, where included, is an estimate of systematic effects.
  In the other cases the systematic errors were estimated to be 
  smaller than the statistical ones.}
\label{results_h}
\end{center}
\end{table}

We remark that in the Higgs phase it was not possible to obtain any
reliable information on excited states. 
In the case of the scalar correlator,
the mass of the first excited state appears to be
about an order of magnitude larger than that of the ground state.
However, the operators of all blocking levels have more than 95\% 
projection onto the
ground state, and correspondingly almost no overlap with the first excitation 
which is hence rather unreliable.
In the case of the field strength correlator the effective masses
computed from ${G}_{F_2}(T)$ do
not show a plateau, and correspondingly no fits are possible.  

The results for the mass in the confinement phase are 
shown in the second block in Table \ref{results_h}. 
Here, we have performed an explicit
check for finite size effects for every value of $\beta_G$ that we simulated.
As is apparent from the table, at least 
all ground state masses have reached their infinite 
volume limits.
At this point in parameter space we were also able to extract some estimates
for the first excitations in the scalar and gauge field channels,
which are labelled by $M_{\Phi_2}$ and $M_{F_2}$,
respectively.

Finally, the results for the two lowest states extracted from the 
field strength correlator 
in the pure SU(2) theory are given in the third block 
in Table \ref{results_h}. Note that the numerical values are very close to 
those found in the confinement phase of the Higgs model. 
This is not unexpected due to the by now familiar fact 
\cite{ptw96,ptw97} that
the dynamics of the gauge degrees of freedom in the confinement phase
of the Higgs model is rather insensitive
to the presence of the scalar fields.
The ground state mass for the pure gauge theory has been previously
calculated in \cite{pou97}. In comparing with that work we note that
our result is $\sim 5\%$ lower. We ascribe this difference to 
a better projection achieved by employing the variational technique.

\subsection{\label{cont} Continuum limit}

\begin{table}[htb]
\centering

\begin{tabular}{||l|l|l||l|l|l|l|l||}
\hline\hline
$x=0.0239$ 
&  &  & $\beta_G=9$ & $\beta_G=12$ & $\beta_G=16$ & $\beta_G=\infty$ &
    $\chi^2$/dof  
\\ \hline\hline
\mbox{Higgs} & $\Phi$ & $M_\Phi/g_3^2$ &
0.1221(1) & 0.1431(1) & 0.1634(2) & -- & -- \\ \cline{3-8} 
\mbox{phase}  &  & $M'_\Phi/g_3^2$ &
-0.0297(1) & -0.0259(1) & -0.0227(2) & -0.014(3) & 2.53 \\ \cline{2-8} 
($y=-0.020$)  & $F$    & $M_F/g_3^2$ & 
2.45(5) & 2.51(2) & 2.61(3) & -- & -- \\ \cline{3-8}
              &        & $M'_F/g_3^2$ & 
2.05(5) & 2.06(2) & 2.12(3) & 2.22(8) & 1.08 \\ \hline\hline
\mbox{confinement} & $\Phi$ & $M_{\Phi_1}/g_3^2$ &
0.727(5) & 0.741(2) & 0.748(6) & -- & -- \\ \cline{3-8}
\mbox{phase} & & $M'_{\Phi_1}/g_3^2$ &
0.575(5) & 0.572(2) & 0.562(6) & 0.55(1) & 1.12 \\ \cline{3-8}
($y=0.089$)  & & $M_{\Phi_2}/g_3^2$ &
1.27(1) & 1.31(4) & 1.31(6) & -- & -- \\ \cline{3-8}
                & & $M'_{\Phi_2}/g_3^2$ &
1.12(1) & 1.14(4) & 1.12(6) & 1.16(10) & 0.16 \\ \cline{2-8}
                 & $F$ & $M_{F_1}/g_3^2$ &
1.523(5) & 1.548(6) & 1.59(2) & -- & -- \\ \cline{3-8}
                 &  & $M'_{F_1}/g_3^2$ &
1.118(5) & 1.097(6) & 1.09(2) & 1.04(3) & 0.17 \\ \cline{3-8}
                 &  & $M_{F_2}/g_3^2$ &
2.17(1) & 2.19(5) & 2.23(3) & -- & -- \\ \cline{3-8}
                 &  & $M'_{F_2}/g_3^2$ &
1.76(1) & 1.74(5) & 1.73(3) & 1.69(6) & 0.03 \\ \hline\hline
\mbox{pure SU(2)} & $F$ & $M_{F_1}/g_3^2$ & 
1.54(2) & 1.55(2) & 1.59(1) & -- & -- \\ \cline{3-8}
          &  & $M'_{F_1}/g_3^2$ & 
1.14(2) & 1.10(2) & 1.09(1) & 1.03(4) & 0.2 \\ \cline{3-8}
          &  & $M_{F_2}/g_3^2$ & 
2.16(3) & 2.23(2) & 2.26(2) & -- & -- \\ \cline{3-8}
          &  & $M'_{F_2}/g_3^2$ & 
1.76(3) & 1.78(2) & 1.77(2) & 1.79(6) & 0.67 \\ \hline\hline
\end{tabular}
\vspace*{1mm}

\caption[a]{\protect \it
Masses in continuum units
for the largest volumes in Table~\ref{results_h} 
(see also Figs.~\ref{contb}-\ref{contg}).
The primed variables are 
after the subtraction in eq.~\nr{subtraction} and are supposed to  
have a continuum limit. The linear continuum extrapolation and the 
corresponding $\chi^2/$dof are shown in the two rightmost columns.
\label{tab:cont}}
\end{table}

\begin{figure}[tbp]

\vspace*{-1.6cm}

\centerline{\hspace{-2mm}
\epsfxsize=9.5cm\epsfbox{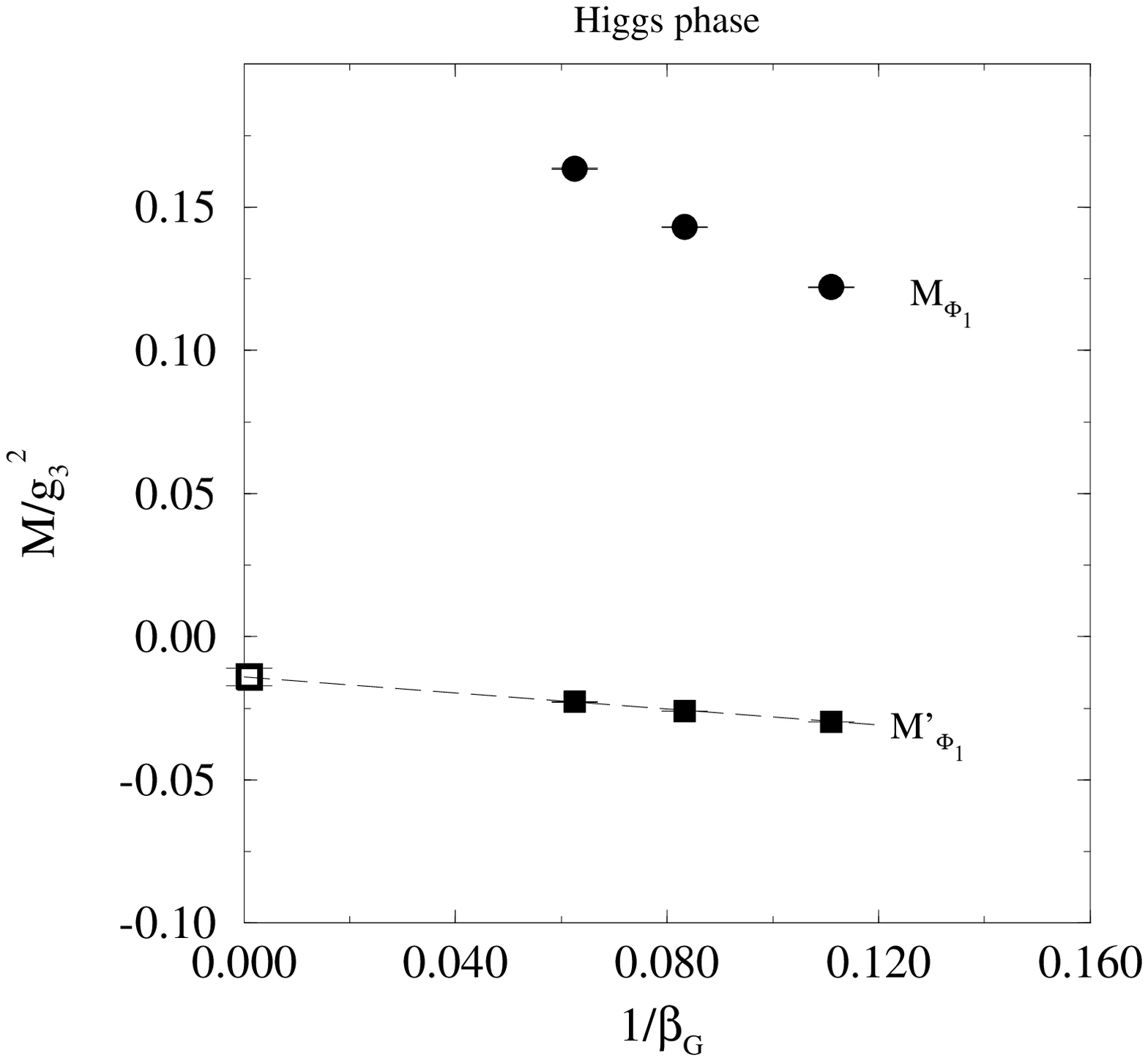}
\hspace{-1.5cm}
\epsfxsize=9.5cm\epsfbox{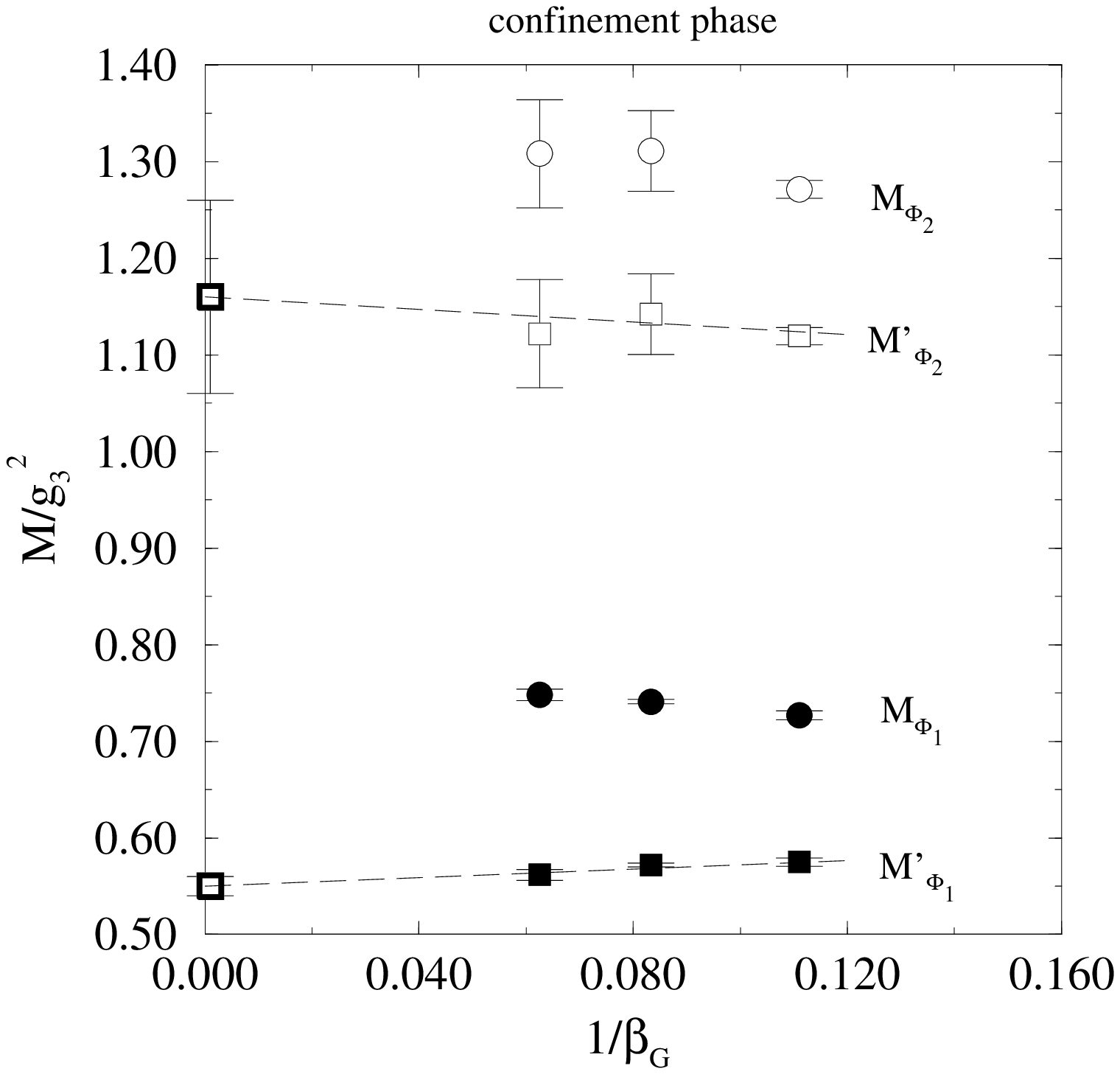}}

\vspace*{-4.5cm}

\caption[]{\label{contb}
{\it The scaling behaviours of $M_{\Phi_1}$, $M_{\Phi_2}$  
     in continuum units.
     Left: Higgs phase.
     Right: confinement phase. 
     The linear continuum extrapolations for the primed quantities 
     are also shown. 
     The logarithmic divergence of $M_{\Phi_1}$ is so weak that 
     it is not clearly  
     visible in the confinement phase figure for the $\beta_G$-values 
     available.}} 
\end{figure}

\begin{figure}[tph]

\vspace*{-1.6cm}

\centerline{\hspace{-2mm}
\epsfxsize=9.5cm\epsfbox{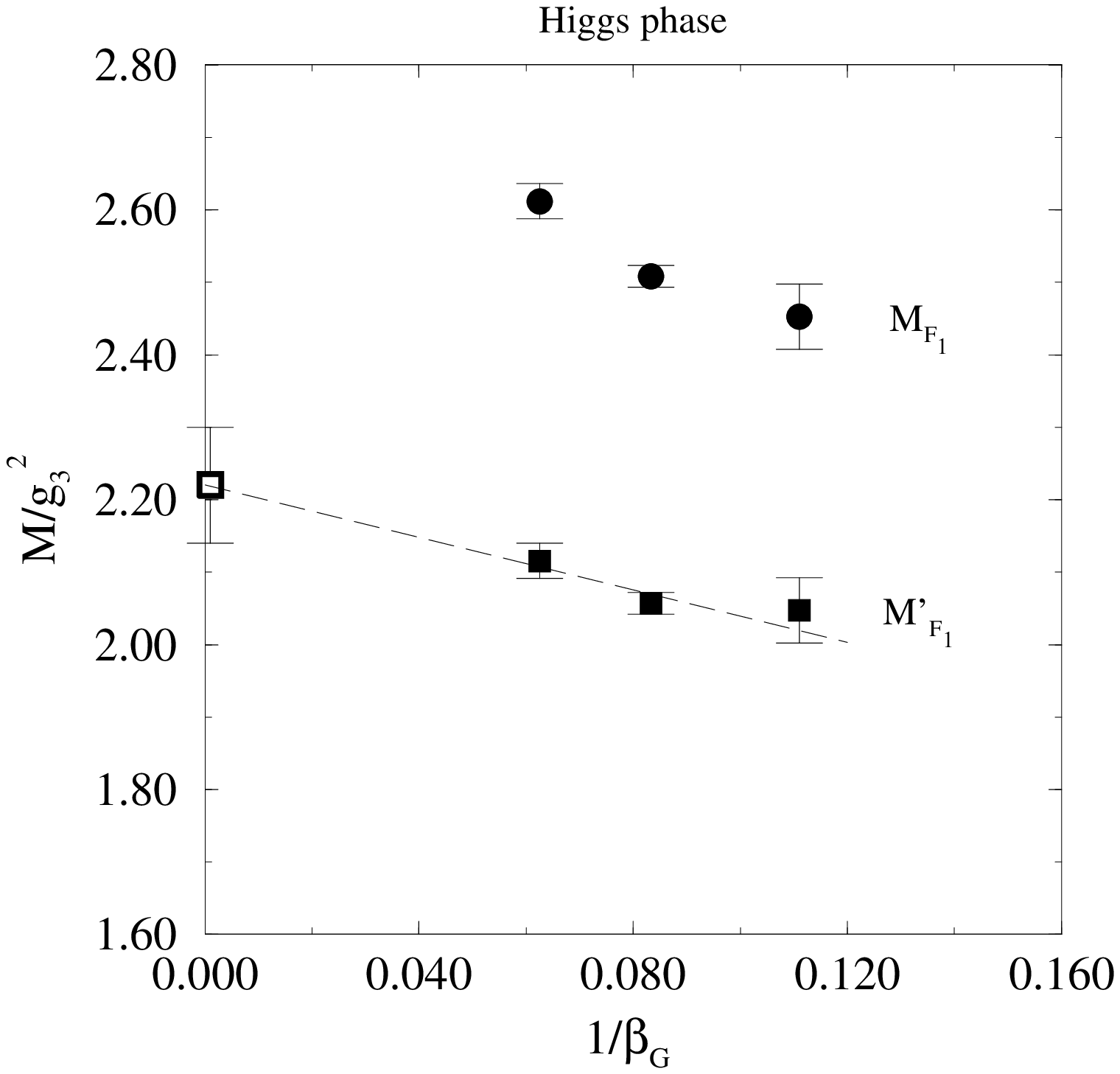}
\hspace{-1.5cm}
\epsfxsize=9.5cm\epsfbox{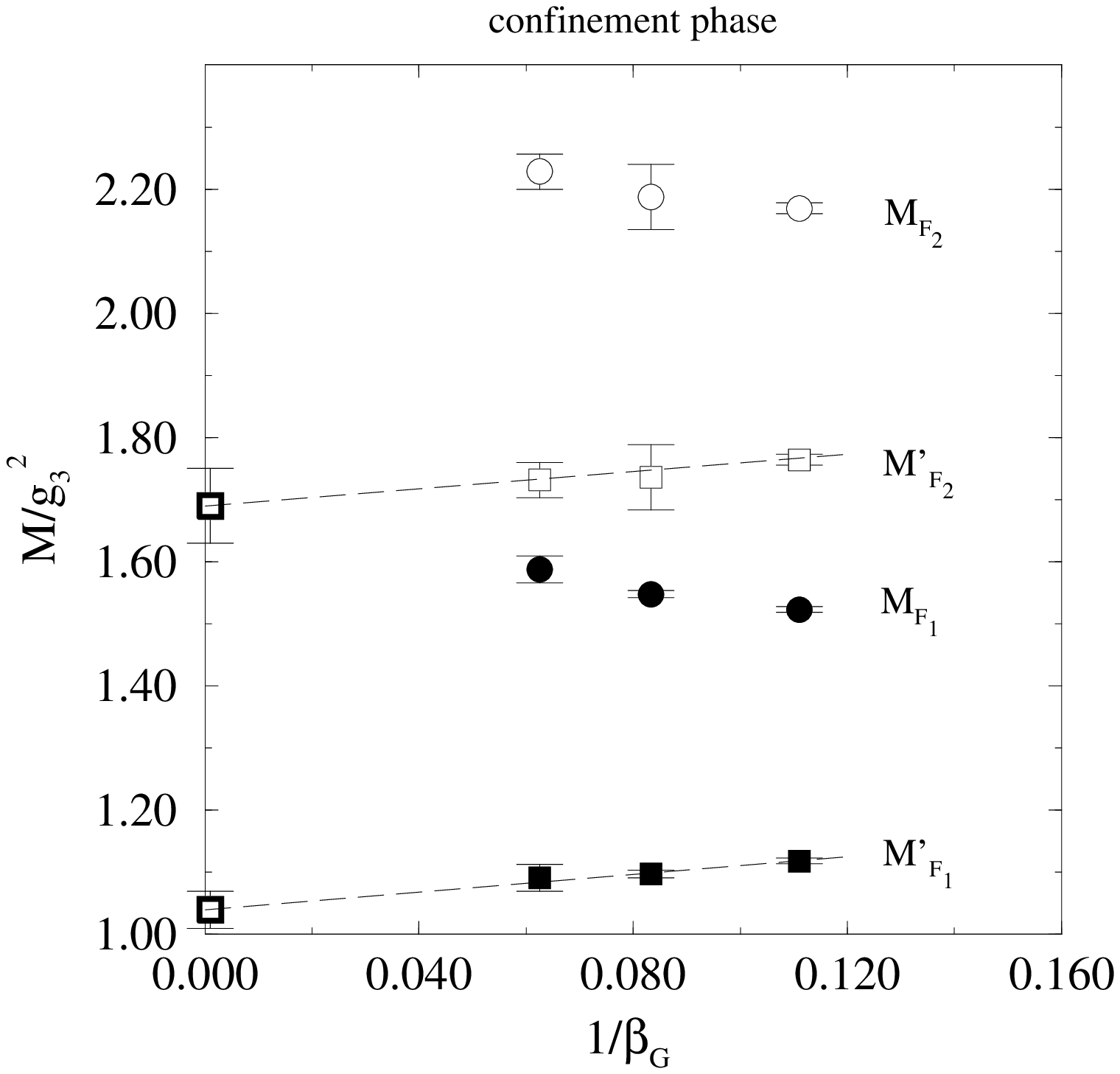}}

\vspace*{-4.5cm}

\caption[a]{\label{conts}
{\it The scaling behaviours of $M_{F_1}$, $M_{F_2}$  
     in continuum units.
     Left: Higgs phase.
     Right: confinement phase. 
     The linear continuum extrapolations for the primed quantities 
     are also shown.}}
\la{beta2}
\end{figure}

\begin{figure}[tb]
 
\vspace*{-1.6cm}
 
\hspace{1cm}
\epsfxsize=9.5cm
\centerline{\epsffile{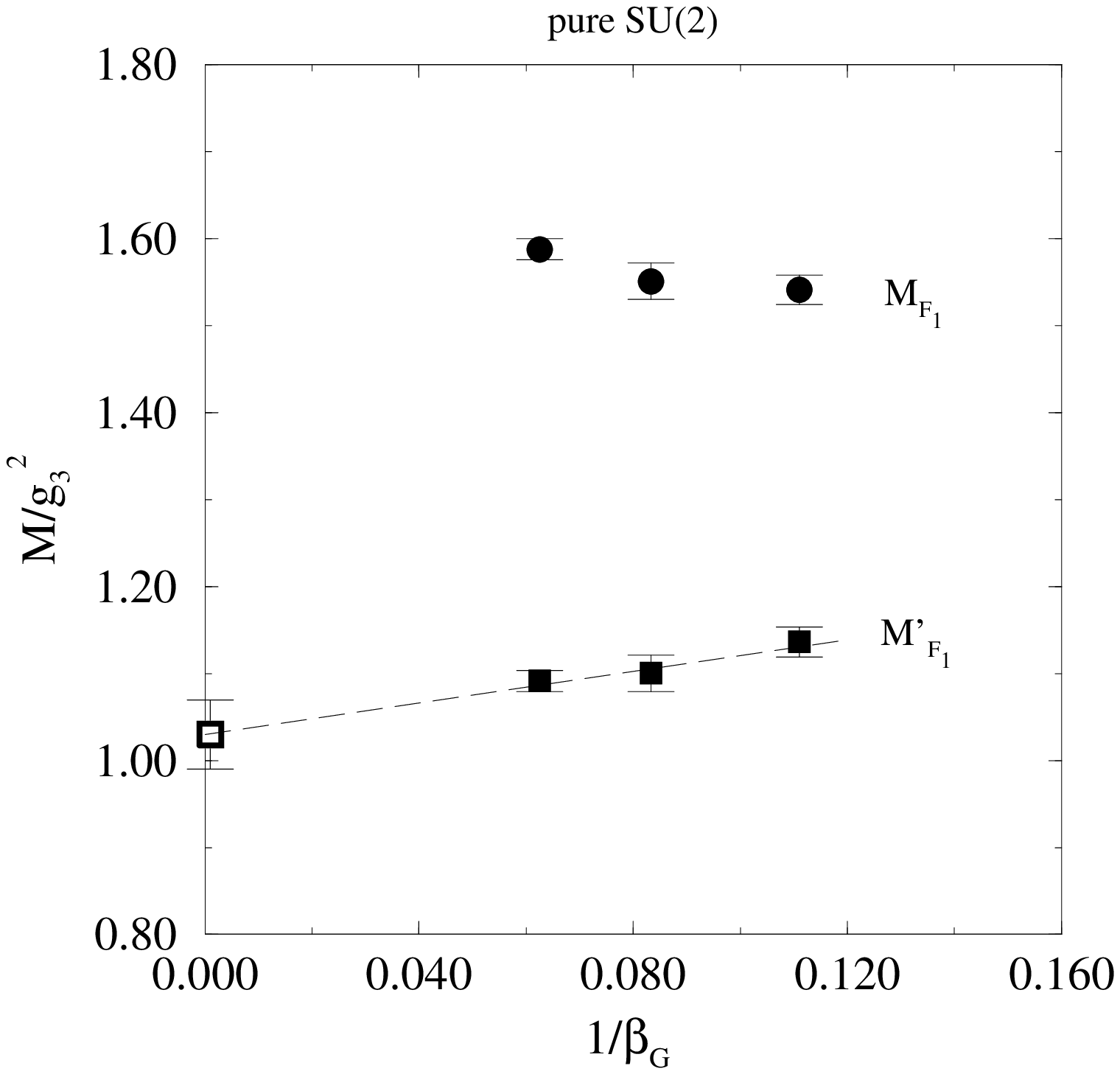}}
 
\vspace*{-4.5cm}
 
\caption[]{\label{contg}
{\it The scaling behaviour of $M_{F_1}$ 
     in the pure SU(2) theory.
     The linear continuum extrapolation for $M'_{F_1}$ is also shown.}} 
\end{figure}

Our next task is to examine the scaling behaviour of our results.
The data are rewritten in continuum units in Table~\ref{tab:cont}.
The scaling of the data with $\beta_G$
is shown in Figs.~\ref{contb}-\ref{contg}.
The observed slight increase of the mass 
values with decreasing lattice spacing
is attributed to the logarithmic divergence 
discussed in Sec.~\ref{div}. Unfortunately,  
the divergence is so weak that in many cases it is not 
possible to clearly observe it numerically, based on our
statistics and $\beta_G$-values. In any case, 
the logarithmic divergence has to be subtracted 
according to eq.~\nr{subtraction}
in order to obtain a finite continuum limit.
The resulting mass values have been denoted with primes.
For the primed quantities we 
observe rather good scaling consistent with linear  
$\sim O(a)$ corrections familiar from calculations of the physical particle
spectrum of the theory \cite{ptw96}. (Let us note that for some observables
the ${\cal O}(a)$ errors could be removed analytically~\cite{moore}, 
but such a computation has not been made for the mass values
obtained from the composite operators in eqs.~\nr{GPhi}, \nr{GA}.)
The outcome of a linear extrapolation of the mass parameters 
in $1/\beta_G$ to $\beta_G=\infty$
is given in Table \ref{tab:cont} and constitutes our final result.

%

In the Higgs phase one can now compare these results with perturbation 
theory. The tree-level perturbative $W$ mass at this 
point is $m_W=1.44 g_3^2$. 
According to eqs.~\nr{subtraction}, \nr{pertGPhi}, 
the 1-loop perturbative scalar mass parameter is thus
\be
{M'_\Phi}^{\rm pert}= -g_3^2\frac{3}{16\pi}
\ln\left(\frac{m_W}{g_3^2}\right)\sim -0.022 g_3^2, \la{pertMPhi'}
\ee
which agrees within $\sim 35\%$ with the full value in Table \ref{tab:cont}.
It should be noted that a large logarithmically divergent 1-loop part 
has been subtracted in $M'_{\Phi}$, so that this agreement is in fact  
quite good. For the vector mass $M'_F$ the order of magnitude 
expectation from the perturbative considerations 
in Sec.~\ref{Hphase} was $M'_F\sim m_W$.
Indeed we observe agreement in order of magnitude, but quantitatively there 
is a discrepancy. 

\section{Discussion}

Let us now discuss the physical
significance our results may have 
for the various topics outlined
in the introduction. 
We begin with the application which has immediate physical meaning.

\subsection{The Debye mass}

An important concept in the phenomenology of high temperature
QCD is the static electric screening mass, or the Debye mass $m_D$. 
However, it is a non-perturbative quantity
beyond leading order~\cite{rebhan}: the expression 
can be written as
\be
m_D = m_D^\rmi{LO}+{Ng_3^2\over4\pi}\ln{m_D^\rmi{LO}\over  
g_3^2} +  
c_N g_3^2 + {\cal O}(g^3T),
\label{md4d}
\ee
where $m_D^\rmi{LO}=(N/3+N_f/6)^{1/2}gT$
and $N_f$ is the number of flavours.   
The logarithmic part of the ${\cal O}(g^2)$ correction 
can be extracted perturbatively \cite{rebhan}, 
but $c_N$ and the higher terms are non-perturbative. 
To allow for a lattice determination, a non-perturbative
definition was formulated in~\cite{ay}, employing
the dimensionally reduced effective theory of eq.~\nr{adact}, 
and the further reduction into the pure 3d SU(N) theory, 
discussed in Sec.~3.3. The statement is that the Debye 
mass can be determined from the exponential fall-off
of the operator odd in $A_0^a$ which gives the lowest
mass value in the theory of eq.~\nr{adact}; 
or, in terms of the pure SU(2) theory,  
from the exponential fall-off of an adjoint Wilson line
with appropriately chosen adjoint charged operators at the ends 
such that $M$ obtains its lowest value. The latter
is precisely the measurement we have made. 

In~\cite{ay} it was further proposed that one could 
measure the constant $M$ from the perimeter law of large
adjoint charge Wilson loops instead of a single Wilson line, 
employing essentially eq.~\nr{asymp}.  
As we have discussed, this measurement is in practice
much more difficult than a straight Wilson line measurement:
so far, it has not been possible to see 
the saturation of the static potential on the lattice
in the adjoint case. 

Identifying now $m_D$ with $M_{HL}$ in Sec.~3.3, 
it follows from eqs.~\nr{mHLmH}, \nr{MH}, \nr{md4d} that
\be
c_N =\frac{M_{F_1}}{g_3^2}+             
\frac{N}{8\pi}\left[\ln\frac{N^2}{2\beta_G^2}-1\right]=
\frac{M'_{F_1}}{g_3^2}+\frac{N}{8\pi}(\ln 4-1).  
\ee
Using the value $M'_{F_1}/g_3^2=1.03(4)$ from Table~\ref{tab:cont}, 
we thus obtain $c_2=1.06(4)$.
 
The constant $c_N$ has previously been determined directly
from the effective theory in eq.~\nr{adact} for $N=2,3$
in~\cite{adjoint2}, where the physical consequences 
of the relatively large                             
non-perturbative correction                         
were discussed, as well.                            
In~\cite{adjoint2},                                 
the result for $N=2$ was found to be $c_2=1.58(20)$. 
This is consistent with our result within $\sim (2\ldots3)\sigma$, 
but our errorbars 
are much smaller.

\subsection{Constituent masses}

In \cite{bp96} it was suggested that the non-local operators
in eqs.~\nr{GPhi}, \nr{GA} would give 
a gauge-invariant handle on the masses of the light constituents forming
the bound states in the confinement phase.
The constituent masses are relevant for the 
bound state model of Ref.~\cite{do95}, as well. In \cite{bp96}
it was further argued that the constituent masses should be 
consistent with the masses determined from the 
exponential fall-off of propagators in a fixed Landau gauge \cite{kar}.
This would then also correspond to the mass value
obtained from gap equations in \cite{bp94}.

Let us first recall that within a constituent model, one would write
the physical bound state mass $M_{HL}$ as $M_{HL}=M_H + M_L + E_B$, 
where $M_H$ and $M_L$ are 
the masses of the static source and the dynamical charge,
respectively, and $E_B$ stands for the binding energy. Then in accordance 
with eq.~\nr{mHLmH}, 
the mass parameter extracted from our correlators may be written as
\beq \label{hl}
M=M_{HL}-M_H=M_{L}+E_{B}.
\eeq
This still leaves the definitions of $M_L$ and $E_B$ open.
In particular, 
it is important to realize that there is a divergence in $M$ 
and thus in $M_L$ or $E_B$. The precise way to handle this divergence
cannot be decided
without a further specification of the constituent model. 

In \cite{bp96} it was suggested that a convenient way of 
circumventing the problem of the divergence and thus of defining 
a finite mass value $M''$ 
(which should approximate $M_L$ in some constituent models),         
would be to write $M=M''+\delta M$ with a constant $\delta M$, and   
to require that in the Higgs phase, $M''$ corresponds
to the physical Higgs or W mass. 
This of course is only possible if the so defined $M''$ has 
the same parametric dependence on, e.g., the continuum parameter $y$
as the physical masses 
in the Higgs phase. If such a matching works, 
then $M''$ will differ by a finite constant
from the finite mass $M'$ defined in eq.~\nr{subtraction}.

\begin{figure}[tb]
 
\vspace*{-1.6cm}
 
\centerline{\hspace{-2mm}
\epsfxsize=9.5cm\epsfbox{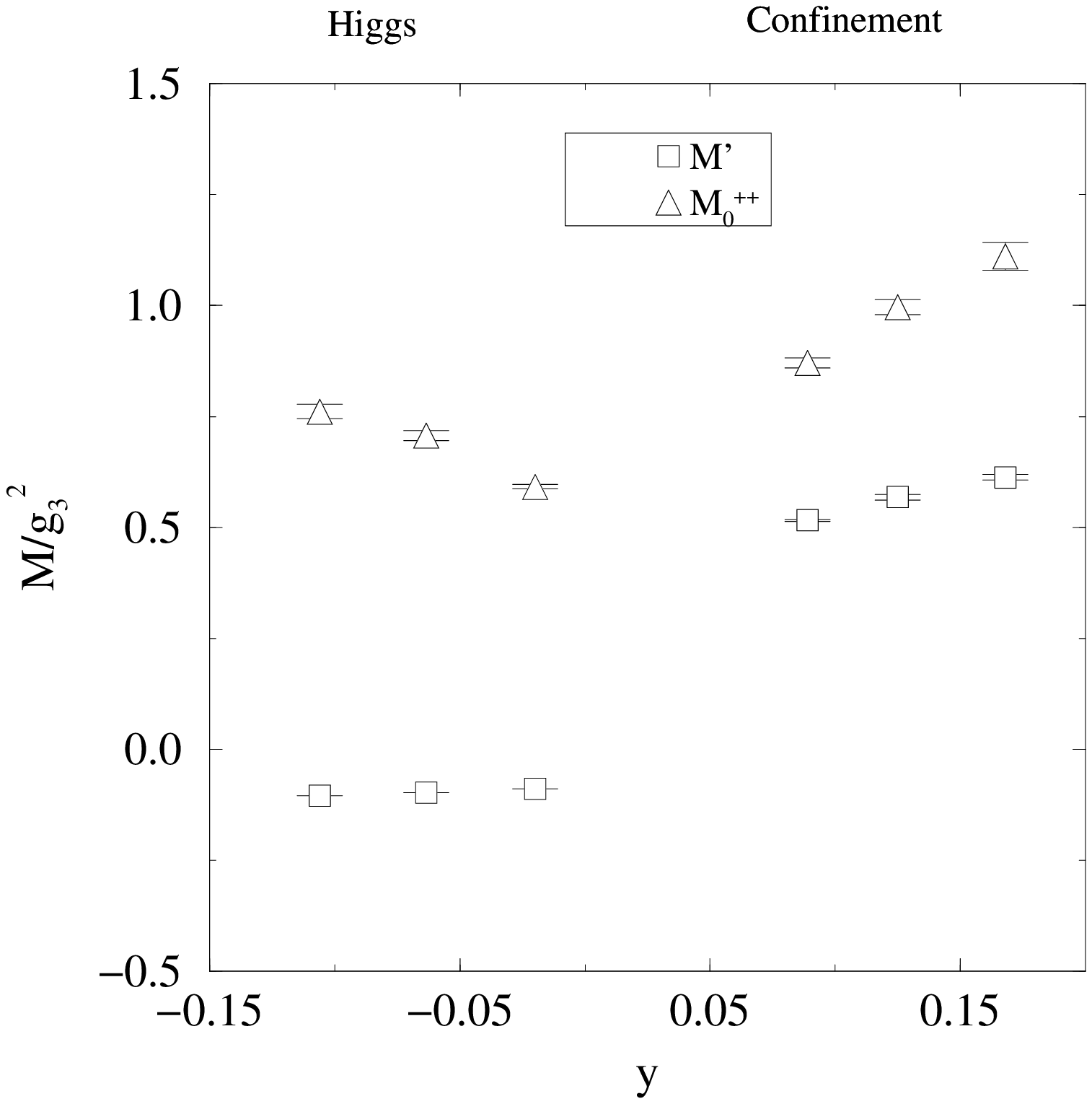}
\hspace{-1.5cm}
\epsfxsize=9.5cm\epsfbox{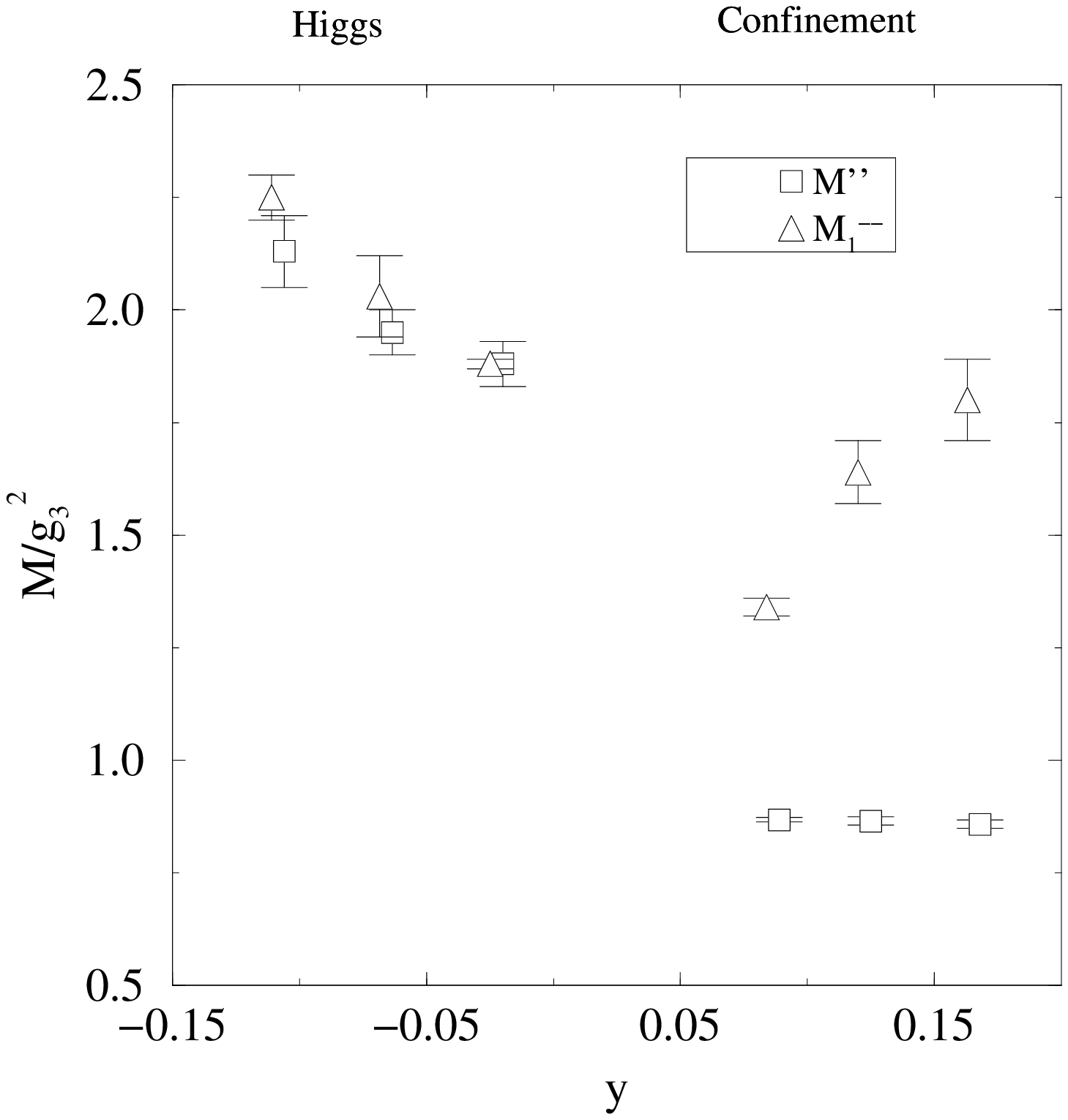}}
 
\vspace*{-4.5cm}
 
\caption[]{\label{mkappa} 
{\it The functional dependence on $y$ at $\beta_G=9$.
     Left: the lowest physical $0^{++}$ state 
     (the Higgs mass in the Higgs phase) and
     the mass $M'$ extracted from
     $G_{\Phi_1}(T)$. Right: the lowest
     physical $1^{--}$ state (the W-boson in the Higgs phase) and
     the mass $M''$ extracted from $G_{F_1}(T)$. 
     The central values of $M''$ have been normalized
     by requiring $M''=M_{1^{--}}$ at $y=-0.02$.}}
\end{figure}

The $y$-dependence of the mass parameters extracted from the 
field correlators
is shown in Fig.~\ref{mkappa} for $\beta_G=9$.
Clearly, the matching procedure 
does not work for the scalar correlator  
since the Higgs phase result 
does not behave as the Higgs mass with respect to $y$. 
The reason is that the 
signal in the Higgs phase is dominated by the cut-off
effect found in perturbation theory, eq.~(\ref{pertGPhi}). 
For the vector correlator, on the other
hand, the parametric dependence of $M''$ in the Higgs phase 
is found to be close to that 
of the W mass 
obtained from the $1^{--}$ operator
$V^a_{i}(x) \sim \tr \left          
(\sigma^a \phi^{\dag}(x) U_i(x) \phi(x+\hat{i})\right)$, 
and such a matching can be applied. 
Setting then
\beq
M''\equiv M'_{F_1}  + A g_3^2\;,
\eeq
A is determined from
\beq
M''(y=-0.020)\equiv m_W(y=-0.020)=1.91 g_3^2 
\eeq
to be $A=-0.31g_3^2$. 
The $y$-dependence of the $M''$ thus obtained 
is shown in Fig.\ref{mkappa} (for $\beta_G=9$), together  
with that of $M_{1^{--}}$.
We observe that across the phase transition, $M''$ jumps to
a smaller value $M''(y=0.089)\approx 0.73 g_3^2$
which then remains fairly constant, in contrast to 
$M_{1^{--}}$. 
This functional behaviour is similar to that of the
W propagator mass in Landau gauge~\cite{kar}, 
as proposed in~\cite{bp96}.   

The numerical value of $M''$ 
in the confinement phase
may now be compared 
with those of other mass parameters
extracted from lattice simulations. 
The lightest 
$0^{++}$ glueball mass is $M_G=1.60(4)g_3^2$ \cite{ptw96};
the fall-off of the W-propagator in Landau gauge in the confinement phase 
has been determined to be
$M^W_{LG}=0.35(1) g_3^2$ \cite{kar}.
We thus observe that the numerical value of $M''$ is  
about half the glueball mass and twice the propagator mass 
in Landau gauge. 
Thus, the conjecture in \cite{bp96}   
according to which $M''\sim M^W_{LG}$, appears
not to be satisfied quantitatively.

\subsection{Screening}

The bound state system of a static source and a dynamical particle
in the fundamental or adjoint representation 
is supposed to correspond to the hadronized
final state of the respective static potential 
at infinite distance,
see eq.~(\ref{asymp}). 
This permits, in principle,
a rough estimate of the screening length
$R_s$ where the static potential flattens off, 
given the behaviour of the potential at small
distances. The problem is that for the adjoint case, 
the potential has also 
been measured up to distances where the screening should show
up if this picture is correct \cite{pou97}, but no sign of the 
screening has been seen. Nevertheless, it might be that 
the determination of the potential at large distances is 
less reliable than at small distances \cite{pou97}. 

Assuming the potential indeed to be determined reliably
only at small distances, one can obtain the following estimates
from 
\beq \label{est}
V(R_s)\approx 2M.  
\eeq
Since $M$ is scheme dependent, one has to use consistently
data from a fixed $\beta_G$.
For the adjoint case in pure SU(2), 
our masses are $\sim 5\%$ smaller
than measured in \cite{pou97}, but for an order of magnitude
estimate this does not matter, and one gets
$R_s^{-1}\sim 0.2 g_3^2$ from Fig.~8 in~\cite{pou97}.
For the fundamental case, we may use the fits made
in \cite{do95,laser} to the $\beta_G=12$ data 
of  \cite{ilg}:
\beq \label{pot}
V(R)=C -\frac{3}{8\pi} g_3^2K_0(mR) + \sigma R,
\eeq
where, corresponding to our confinement phase point, 
the extrapolated fit parameters have been estimated as 
$C=0.310 g_3^2$, $m=1.05 g_3^2$ and                       
$\sigma=0.133 g_3^4$~\cite{laser}. Using the $\beta_G=12$ 
mass $M_{\Phi_1}=0.740 g_3^2$
from Table~\ref{tab:cont}, eq.~\nr{est} then gives
$R_s^{-1}\sim 0.1 g_3^2$ for the fundamental case.   
Given the uncertainties, these numbers are not to be trusted
quantitatively, but it is nevertheless 
interesting to observe that both
the adjoint and the fundamental case give a 
screening length of the same order of magnitude, 
and that the corresponding mass
scale $R_s^{-1}$ is quite small compared with all the 
correlator masses. 

\section{Conclusions}

In this paper we have reported on the measurements 
of the expectation values of the operators
in eqs.~\nr{GPhi}, \nr{GA} for Wilson lines with various
lengths $|x-y|$, on 3d lattices with various lattice sizes and lattice 
spacings. This allows the extraction of mass signals related
to the exponential fall-off of the expectation 
values, and a study of their scaling behaviour. Subtracting
a perturbatively computable logarithmic divergence, 
we have been able to
extrapolate the non-perturbative constant 
parts to the continuum limit.
Apart from the mass signal in the exponential 
fall-off, we have also measured the asymptotic 
functional forms of the correlation functions, 
verifying that they behave according to what 
is expected for confining theories. 

The mass thus measured from eq.~\nr{GA}
for the pure 3d SU(2) theory allows a determination of 
the leading non-perturbative contribution to the finite
temperature Debye mass $m_D$ with a method
due to Arnold and Yaffe~\cite{ay}. We obtain 
a much more precise estimate for $m_D$  
than has been achieved before: 
$m_D= m_D^\rmi{LO}+{g_3^2/(2\pi)}\ln{(m_D^\rmi{LO}/  
g_3^2)} +  
1.06(4) g_3^2 + {\cal O}(g^3T)$. 
The present
measurement was made for SU(2) QCD. However, 
now that our study has proven the practical feasibility of this
method, the extension to the realistic SU(3) case is 
straightforward.  
The physical significance of the relatively large  
non-perturbative correction                       
has been discussed in~\cite{adjoint2}.            

On the more phenomenological side, the mass
signals measured from eqs.~\nr{GPhi}, \nr{GA} 
are relevant for the composite models proposed to apply
in the (symmetric) confinement phase of the SU(2)+Higgs model. 
This system has been studied 
as a laboratory for understanding confinement. 
The main observation based on our data is that, 
with a definite subtraction procedure, 
the lattice spacing independent part of the  
field strength correlator mass is large: 
it is consistent with about half the glueball mass 
and twice the mass of the W propagator in Landau gauge. 

Finally, we have addressed the question of screening of
the static potential. 
Assuming that the static potential measurements up to date 
are only reliable at small distances, one obtains 
both in the pure SU(2) theory and in the SU(2)+Higgs model the 
rough estimate $1/R_s\sim 0.1\ldots 0.2 g_3^2$ for the screening
length of adjoint and fundamental representation charges. 
It is interesting that the mass scale here is much smaller
than the other mass scales in the system. 

\subsection*{Acknowledgements}

We thank M.~G\"urtler, E.-M.~Ilgenfritz and A.~Schiller 
for providing us with data allowing for a check of our code, 
H.~Wittig for routines and useful advice on fitting, and 
W.~Buchm\"uller, M.G.~Schmidt and M.~Teper 
for interesting discussions.
The calculations were perfomed on a NEC-SX4/32 at the HLRS Universit\"at
Stuttgart.


\begin{thebibliography}{99}

\bibitem{bali}
G.~Bali and K.~Schilling, Phys.~Rev.~D 46 (1992) 2636; 
Phys.~Rev.~D 47 (1993) 661; Int.~J.~Mod.~Phys.~C 4 (1993) 1167;
S.~Booth et al., Nucl.~Phys.~B 394 (1993) 509;
S.~G\"usken, in {\it Proceedings of Lattice '97}, in press [hep-lat/9710075]. 

\bibitem{mic85}
C. Michael, Nucl.~Phys.~B 259 (1985) 58; 
N.A. Campbell, I.H. Jorysz and C. Michael,
Phys.\ Lett.\ B 167 (1986) 91;
I. Jorysz and C. Michael,
Nucl.~Phys.~B 302 (1987) 448;
C. Michael, Nucl.~Phys.~B (Proc.\ Suppl.) 26 (1992) 417.

\bibitem{ilg}
E.-M.~Ilgenfritz, J.~Kripfganz, H.~Perlt and A.~Schiller,
Phys.\ Lett.\ B 356 (1995) 561;
M.~G\"urtler, E.-M.~Ilgenfritz, J.~Kripfganz, H.~Perlt and A.~Schiller,
Nucl.\ Phys.\ B 483 (1997) 383.

\bibitem{pou97}
G.I. Poulis and H.D. Trottier, Phys.~Lett.~B 400 (1997) 358.

\bibitem{bri83}
J. Bricmont and J. Fr\"ohlich, Phys.~Lett.~B 122 (1983) 73.

\bibitem{gromes}
D. Gromes, 
Phys.\ Lett.\ B 115 (1982) 482.

\bibitem{do87}
H.G. Dosch, 
Phys.\ Lett.\ B 190 (1987) 177;
H.G. Dosch and Y.A. Simonov, 
Phys.\ Lett.\ B 205 (1988) 339.

\bibitem{do95}
H.-G.~Dosch, J.~Kripfganz, A.~Laser and M.G.~Schmidt,
Phys.~Lett.~B 365 (1995) 213; HD-THEP-96-53 [hep-ph/9612450]. 

\bibitem{gia}
A.D. Giacomo, E. Meggiolaro and H. Panagopoulos, 
Nucl.\ Phys.\ B 483 (1997) 371;
M. D'Elia, A.D. Giacomo and E. Meggiolaro, 
IFUP-TH 19/97 [hep-lat/9705032].

\bibitem{jamin}
M. Eidem\"uller and M. Jamin, 
HD-THEP-97-49 [hep-ph/9709419].

\bibitem{bbv}
G.S. Bali, N. Brambilla and A. Vairo, 
HD-THEP-97-35 [hep-lat/9709079].

\bibitem{qcd3}
J. Ambj{\o}rn, P. Olesen and C. Peterson, 
Nucl.\ Phys.\ B 240 (1984) 533;
D.G. Caldi and T. Sterling, 
Phys.\ Rev.\ Lett.\ 60 (1988) 2454;
R.D. Mawhinney, Phys.\ Rev.\ D 41 (1990) 3209;
H.D. Trottier and R.M. Woloshyn, 
Phys.\ Rev.\ D 48 (1993) 2290;
M. Teper, Phys.~Lett.~B 289 (1992) 115; Phys.~Lett.~B 311 (1993) 223.

\bibitem{ptw97}
O.~Philipsen, M.~Teper and H.~Wittig,
HD-THEP-97-37 [hep-lat/9709145].

\bibitem{kaj96}
K.~Kajantie, M.~Laine, K.~Rummukainen and M.~Shaposhnikov,
Nucl.\ Phys.\ B 466 (1996) 189;
Nucl.\ Phys.\ B 493 (1997) 387.

\bibitem{ptw96}
O.~Philipsen, M.~Teper and H.~Wittig, Nucl.~Phys.~B 469 (1996) 445.

\bibitem{isthere?}
K.~Kajantie, M.~Laine, K.~Rummukainen and M.~Shaposhnikov,
Phys.\ Rev.\ Lett.\ 77 (1996) 2887.

\bibitem{old}
P. Ginsparg,
Nucl.\ Phys.\ B 170 (1980) 388;
T. Appelquist and R. Pisarski,
Phys.\ Rev.\ D 23 (1981) 2305.

\bibitem{dr}
K.\ Kajantie, M.\ Laine, K.\ Rummukainen and M.\ Shaposhnikov,
Nucl.\ Phys.\ B 458 (1996) 90; 
E. Braaten and A. Nieto, 
Phys.\ Rev.\ D 53 (1996) 3421;
A.~Jakov\'ac and A.~Patk\'os, 
Nucl.\ Phys.\ B 494 (1997) 54.

\bibitem{ei88}
E. Eichten, Nucl.~Phys.~B (Proc.~Suppl.) 4 (1988) 170.

\bibitem{bp96}
W. Buchm\"uller and O. Philipsen, Phys.~Lett.~B 397 (1997) 112.

\bibitem{kar}
F.~Karsch, T.~Neuhaus, A.~Patk\'os and J.~Rank, 
Nucl.~Phys.~B 474 (1996) 217.

\bibitem{ay}
P. Arnold and L. Yaffe, Phys.~Rev.~D 52 (1995) 7208.

\bibitem{ev86}
H.G.~Evertz, V.~Gr\"osch, J.~Jers\'ak, H.A.~Kastrup, T.~Neuhaus,
D.P.~Landau and J.-L.~Xu, Phys.~Lett.~B 175 (1986) 335.

\bibitem{leip}
M. G\"urtler, E.M. Ilgenfritz, A. Schiller and C. Strecha, in
{\em Proceedings of Lattice'97}, in press [hep-lat/9709020].

\bibitem{fkrs95}
K. Farakos, K. Kajantie, K. Rummukainen and M. Shaposhnikov, 
Nucl.\ Phys.\ B 442 (1995) 317;
M.~Laine, Nucl.~Phys.~B 451 (1995) 484; 
M. Laine and A. Rajantie,    
Nucl.\ Phys.\ B, in press [hep-lat/9705003].  

\bibitem{cum}
N.G. van Kampen, 
Phys.\ Rep.\ 24 (1976) 171.

\bibitem{rebhan} A.K. Rebhan,
Phys. Rev. D 48 (1993) R3967;
Nucl. Phys. B 430 (1994) 319.

\bibitem{adjoint2}  
K. Kajantie, M. Laine, K. Rummukainen and M. Shaposhnikov, 
Nucl. Phys. B, in press [hep-ph/9704416];
K. Kajantie, M. Laine, J. Peisa, A. Rajantie, 
K. Rummukainen and M. Shaposhnikov, 
Phys.\ Rev.\ Lett.\ 79 (1997) 3130 [hep-ph/9708207].

\bibitem{fab}
K.~Fabricius and O.~Haan, Phys.~Lett.~B 143 (1984) 459.

\bibitem{ken}
A.D.~Kennedy and B.J.~Pendleton, Phys.~Lett.~B 156 (1985) 393.

\bibitem{bunk}
B.~Bunk, Nucl.~Phys.~B (Proc.~Suppl.) 42 (1995) 566.

\bibitem{tep87}
M.~Teper, Phys.~Lett.~B 183 (1987) 345.

\bibitem{moore}
G.D. Moore,
Nucl. Phys. B 493 (1997) 439;
McGill-97-23 [hep-lat/9709053].

\bibitem{bp94}
W. Buchm\"uller and O. Philipsen, Nucl.~Phys.~B 443 (1995) 47.

\bibitem{laser}
A. Laser, PhD thesis, Heidelberg 1996 (unpublished).

\end{thebibliography}
\end{document}